 \documentclass[structabstract]{aa}
\usepackage{graphicx}
\usepackage{txfonts}
\usepackage{natbib}
\usepackage{color}

\begin{document}

   \title{Modeling optical and UV polarization of AGNs}

   \subtitle{III. From uniform-density to clumpy regions}

   \author{F.~Marin\inst{1}\thanks{\email{frederic.marin@asu.cas.cz}}
     \and R.~W. Goosmann\inst{2}
     \and C.~M. Gaskell\inst{3}
     }

   \institute{Astronomical Institute of the Academy of Sciences, 
     Bo{\v c}n\'{\i} II 1401, CZ-14100 Prague, Czech Republic
     \and Observatoire Astronomique de Strasbourg, Universit\'e de Strasbourg,
     CNRS, UMR 7550, 11 rue de l'Universit\'e, 67000 Strasbourg, France
     \and Department of Astronomy and Astrophysics, University of California at Santa Cruz, California 95064, USA}

  \date{Received 8 January 2015; Accepted 17 March 2015}

  \abstract {A growing body of evidence suggests that part of, if not
    all, scattering regions of active galactic nuclei (AGNs) are
    clumpy. The inner AGN components cannot be spatially resolved with
    current instruments and must be studied by numerical simulations
    of observed spectroscopy and polarization data.} 
    {We run radiative transfer models in the optical/UV for a variety of AGN
    reprocessing regions with different distributions of clumpy
    scattering media. We obtain geometry-sensitive polarization
    spectra and images to improve our previous AGN models and their
    comparison with the observations.} 
    {We use the latest public version 1.2 of the Monte Carlo code STOKES presented in
    the first two papers of this series to model AGN reprocessing
    regions of increasing morphological complexity. We replace
    previously uniform-density media with up to thousands of constant-density
    clumps. We couple a continuum source to fragmented equatorial
    scattering regions, polar outflows, and toroidal, obscuring dust
    regions and investigate a wide range of geometries. We also
    consider different levels of fragmentation in each scattering
    region to evaluate importance of fragmentation for the net 
    polarization of the AGN.} 
    {In comparison with uniform-density models, equatorial distributions of 
    gas and dust clouds result in grayer spectra, and show a decrease of the 
    net polarization percentage at all lines of sight. The resulting polarization
    position angle depends on the morphology of the clumpy structure,
    with extended tori favoring parallel polarization while compact
    tori produce orthogonal polarization position angles. In the case
    of polar scattering regions, fragmentation increases the net
    polarization unless the cloud filling factor is small. A complete
    AGN model constructed from the individual, fragmented regions can
    produce low polarization percentages ($<$ 2~\%), with a parallel
    polarization angle for observer inclinations up to
    70$^\circ$ for a torus half opening angle of 60$^\circ$. For type-2 
    viewing angles the polarization switches to perpendicular and rises 
    to $\sim$ 50~\%.} 
    {Our modeling shows that the introduction of fragmented dusty tori
    significantly alters the resulting net polarization of an AGN;
    Comparison of our models to polarization observations of large AGN
    samples greatly favors geometrically compact clumpy tori over 
    extended ones.}
   

\maketitle


\section{Introduction}

Fragmentation is a ubiquitous phenomenon in astrophysics. Theories
exploring the stability of uniform-density, self-gravitating media predict
fragmentation if the initial amount of matter is spread over large
distances \citep{Jeans1902,Hunter1962,Arny1966}. However, observations of
fragmented systems in external galaxies are complicated by the small size 
of the resulting clumps. Current instruments are unable to fully resolve 
the detailed morphology of distant yet luminous systems such as AGN. 

Optical/UV observations of the narrow-line region (NLR) in
NGC~1068, obtained by \citet{Evans1991}, and later by
\citet{Capetti1995}, revealed the presence of several knots of
different luminosity in the outflowing gas. These authors identified
these clumps with inhomogeneities in the medium. Similar results are
found in other studies at the same \citep{Capetti1997} or longer
wavelengths \citep{Packham1997}, strengthening the idea that the
outflows of NGC~1068 are not continuous. More recently,
\citet{Aalto2012}, using the Plateau de Bure millimeter
interferometer, found that the large-scale molecular wind of the
ultraluminous infrared galaxy Mrk~231 is likely to consists of dense,
clumpy gas clouds with high abundances of hydrogen cyanide (HCN). They
attribute these HCN properties to shock-driven compression and
fragmentation in the outflows. 

There is less direct evidence for the well-known ``dusty torus'', the
supposed bulk circumnuclear medium shielding the inner AGN along the
equatorial plane. The torus size invoked by the unified model
\citep{Antonucci1993}, typically from 0.1 to 100 parsecs, is inconsistent 
with self-gravitational stability. \citet{Krolik1988} and \citet{Pier1992} 
argued on theoretical grounds that AGN tori are likely to consist of 
individual, optically thick, molecular clumps in collision-free orbits 
that are sustaining the vertical torus height required by observations. 
Self-shielded from the full continuum, the gas/dust clouds would survive 
the few hundred Kelvin temperatures in the inner toroidal regions 
\citep{Li2007}. As noted by \citet{Nenkova2002}, the 10~$\mu$m silicate 
feature is a consequence of the torus being clumpy.

Fragmented dynamical dust distributions in AGNs have started to be
investigated in numerical simulations. Based on the suggestions of
\citet{Pier1992,Pier1993} about the nature of the obscuring dust,
\citet{Nenkova2002} undertook IR modeling of a distribution of clumps
around a central irradiating source. The authors successfully
reproduced the inclination-dependent behavior of the 10~$\mu$m silicate 
feature, the width of the far-IR emission peak, and the 0.01 -- 100~$\mu$m 
spectral energy distribution (SED) of both type-1 and type-2 objects. 
Similar work in the far-IR -- UV wavelengths 
\citep{Nenkova2008a,Nenkova2008b,Nenkova2010,Schartmann2008,Honig2010,Heymann2012,Schartmann2014}, 
and in the X-ray band \citep{Risaliti2002,Liu2014} was conducted to reproduce 
and/or study the spectroscopic behavior of fragmented tori.

The interferometric observations of the type-2 Circinus galaxy undertaken 
by \citet{Tristram2007} with the mid-infrared (MIDI) instrument at 
the Very Large Telescope, strongly support the presence of a clumpy
toroidal structure. The data are inconsistent with a smooth density
distribution and indicate clumpy or filamentary structures. In a
fragmented medium, the low-density gaps between the clouds allow
radiation to propagate between optically thick clumps so that
obscuration along the equatorial plane is less effective. 
In addition to this, the equilibrium between cloud mergers 
and tidal shearing keeps the covering factor of the clump distribution 
close to unity in the equatorial direction \citep{Krolik1988}. 
Since the scattering probability increases with
the filling factor of a fragmented region, an impact on the
polarization properties of the escaping radiation is expected. 
UV/optical spectropolarimetry thus becomes an ideal counterpart to 
mid-infrared interferometry to explore the morphology of AGNs.

In this paper, the third of our series (Paper~I: \citealt{Goosmann2007},
Paper~II: \citealt{Marin2012}), we investigate, for the first time,
the UV/optical polarization of AGN with fragmented reprocessing
regions. We compare uniform-density models of dusty tori, outflowing
winds and accretion flows to their fragmented counterparts and build
an AGN model made of more than four thousands reprocessing clumps. We
also explore different configurations of the clumpy AGN model to test
the impact of several morphological parameters and analyze our results
with respect to polarimetric observations.

The remainder of the paper is organized as follows: in
Sect.~\ref{Clumpy:Impact} we briefly summarize our previous modeling
work on AGN with {\sc stokes} and study polarization signatures of
individual, clumpy reprocessing regions. A three-component model
approximating the unified scheme of AGN is explored in
Sect.~\ref{Clumpy:AGN}. In Sect.~\ref{Clumpy:Discussion} we discuss
our results and relate them to past spectropolarimetric observations
before drawing our conclusions in Sect.~\ref{Clumpy:Conclusion}.

\section{Polarization signatures from clumpy regions}
\label{Clumpy:Impact}

In the context of multiple reprocessing between a large variety of
emitting, scattering and absorbing regions, it is important to first
test how an individual fragmented region differs from a
uniform-density one. Because polarimetry is sensitive to the 
geometry of the reprocessing medium, a clumpy distribution of spheres
is likely to alter the net polarization degree in a different way than 
uniform-density regions. We thus reanalyze the individual reprocessing 
regions presented in Papers~I~and~II after introducing clumpiness. 

For the remainder of this paper, we loosely define a {\it type-1} 
AGN orientation as an inclination where the observer's line of sight toward 
the central source that does not intercept the torus boundaries\footnote{
This definition assumes a sharp-edge torus in angular direction; it is a 
basic simplification as real tori are likely to have soft-edges 
\citep[see, e.g.,][]{Alonso2003}.}. In this case, the viewing angle, $i$, 
is smaller than the half-opening angle of the toroidal structure, with both 
angles being measured from the symmetry axis. Otherwise, the object is a 
{\it type-2} AGN (such as in Fig.~\ref{Fig:Scheme}).

Polarization of light will be described by its polarization degree,
$P$, ranging from 0~\% (unpolarized) to 100~\% (fully polarized), and
by its polarization position angle $\gamma$. If the $\vec E$-vector of
the radiation is aligned with the projected torus axis, the
polarization is called \textit{parallel} ($\gamma$ = 90$^\circ$). When
$\gamma$ = 0$^\circ$, polarization is said to be \textit{perpendicular}. 
A convention, already used in Paper~I and II,
identifies parallel and perpendicular polarization by the sign of $P$:
a negative value of $P$ stands for parallel polarization, a positive
$P$ for perpendicular one. Plotting $P$ over inclination thus allows
us to easily identify the observed polarization dichotomy in thermal
AGNs (type-1 AGNs being generally polarized parallel to the axis of
the torus while type-2 AGNs being polarized perpendicularly). Finally,
our computations include both linear and circular polarization and the
sum of the two is shown when plotting $P$. However, circular
polarization has nearly no impact on the total polarization, being at
least a hundred times lower than linear polarization.

\subsection{A brief overview of {\sc stokes}}

{\sc stokes} is a public Monte Carlo code that simulates 
the radiative transfer between radiatively coupled emitting and
reprocessing media and computes the polarization properties as a
function of the viewing direction from the near-infrared to the hard
X-ray band. The latest release, {\sc stokes 1.2}, can be downloaded
from the internet\footnote{http://www.stokes-program.info/}. In
\citet{Goosmann2007} [Paper~1], we presented the first version of the
code and investigated how the morphology of individual AGN
reprocessing regions influences the resulting polarization. In
\citet{Marin2012} [Paper~II], we upgraded the computational speed of
the code and added an imaging routine. We developed an AGN model
composed of three different scattering media and explored the
parameter space that could reproduce the spectropolarimetric behavior
of different AGN types, such as Seyfert galaxies and quasars, but also
``naked'', ``bare'', FR-I, LINERs. Since then, our code has been used
to explore a large range of AGN properties (see \citealt{Marin2014b}
for a list of examples).

The fast performance the code allows us to significantly increase the
number of scattering regions. Multiple scattering is a prerequisite in such
models since individual clouds are radiatively coupled.
The version 1.2 of {\sc stokes} is able to process a model with more
than 10~000 individual scattering clouds and to obtain good-quality
spectra by sampling 5 $\times$ 10$^8$ photons or less in a timescale of
a day (depending on the presence of absorbers and the required precision).

\subsection{Characterizing a fragmented medium}

\begin{figure}
   \centering
   \includegraphics[trim = 0mm 110mm 0mm 60mm, clip, width=9cm]{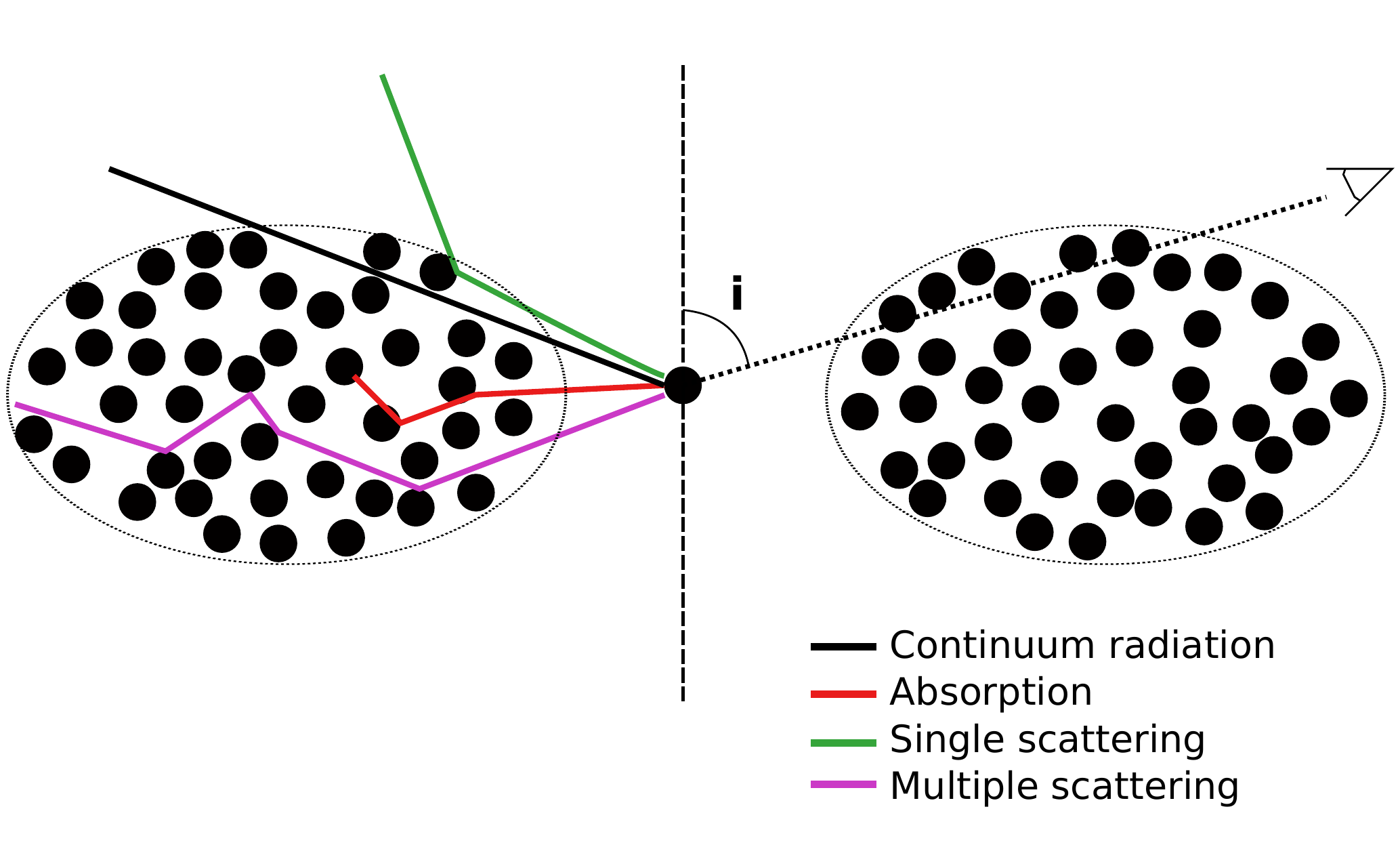}
      \caption{Schematic view of a fragmented region. Clumps are shielding each other
	       from the central source, but gaps between the spheres allow part of radiation
	       to escape from the model, even along equatorial directions where obscuration
	       is most efficient (in the case of circumnuclear dust).}
     \label{Fig:Scheme}%
   \end{figure}

The overall picture of AGN becomes more complicated when replacing
uniform-density components by their clumpy counterparts. The radiative
transfer calculations take more time as the number of scattering
increases. Complex radiative interactions arise between consecutive
clouds and alter the resulting spectra and polarization (see
Fig.~\ref{Fig:Scheme}) at a given viewing angle.

%

   \begin{figure*}
   \centering
   \includegraphics[trim = 0mm 240mm 0mm 0mm, clip, width=18cm]{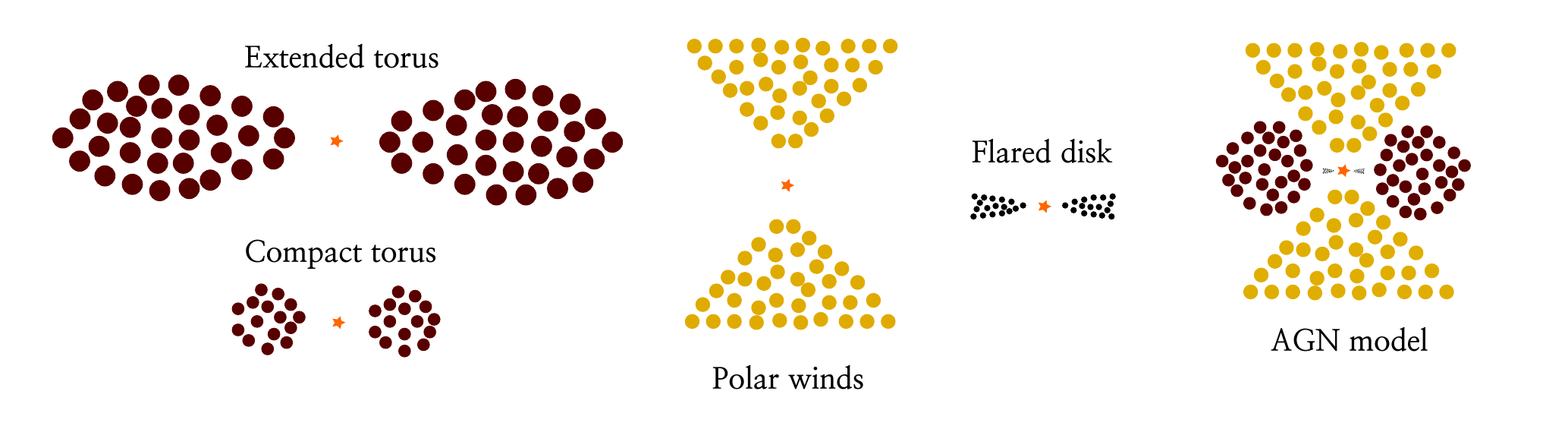}
      \caption{Sketches of the different models investigated in this paper.
	       Tori with cylindrical cross-section, either extended or compact, 
	       are explored in Sect.~\ref{Clumpy:Impact:Torus}, polar winds 
	       in Sect.~\ref{Clumpy:Impact:Winds}, and flared disks in 
	       Sect.~\ref{Clumpy:Impact:Flared}. Partially and fully 
	       fragmented AGN models are presented in Sect.~\ref{Clumpy:AGN}.}
     \label{Fig:Models}%
   \end{figure*}
%

To investigate the impact of fragmentation, we first re-compute the polarization 
of isolated, uniform-density scattering regions presented in Paper~II, namely the 
dusty torus, the polar winds and the hot accretion flow. We conserve the model 
geometry but replace the region by clumpy distributions of spheres, such as shown 
in Fig.~\ref{Fig:Models}. These spheres are distributed uniformly inside the 
scattering region without overlapping, leaving gaps between clouds. To characterize 
such media and evaluate how fragmentation affects the polarization, we rely on the 
filling factor $\cal{F}$, ranging from 0 to 1 (uniform-density region). 
$\cal{F}$ is evaluated by summing up the volume of the clumps and dividing the 
total by the volume of the same, unfragmented reprocessing region. 

For the remainder of this paper, several thousand of clumps are modeled as constant 
density spheres of equal radii. Three filling factors are investigated: 0.06 (highly 
fragmented region), 0.13 and 0.25 (moderate fragmentation). In comparison, 
in their X-ray spectral model for clumpy tori, \citet{Liu2014} used a filling 
factor $\le$ 0.1 for their fragmented models. The total optical depth along the 
observer's viewing angle strongly depends on $\cal{F}$: for a model 
with $\cal{F} \sim$~0.06, 1 to 3 clumps can pile-up along a line of sight that crosses 
the cloud distribution, while tens of clumps can hide the irradiating source
in the case of $\cal{F} \sim$~0.25.

~\

In the following sections, we investigate the spectropolarimetric
signature of fragmented dusty tori, polar outflows and hot accretion
flows. The results of each clumpy model will be compared with the
polarization and flux (normalized to the flux of the central source)
of the same, but uniform-density model. The parametrization of those
models is based on the set of constraints derived from Paper~I and
II. For all our models, we define an isotropic, point-like source at
the center of the model emitting an unpolarized spectrum with a
power-law SED\footnote{The shape of the SED is of no importance for 
what we are considering here.} $F_{\rm *}~\propto~\nu^{-\alpha}$ and $\alpha = 1$.

\subsection{Clumpy torus models}
\label{Clumpy:Impact:Torus}

\subsubsection{Extended tori}
\label{Clumpy:Impact:Torus_Extended}

Our first model features an obscuring dusty torus with an elliptical 
cross-section, centered on the source, symmetric to the equatorial plane. 
According to the simple unified scheme \citep{Antonucci1993}, its outer boundary 
can extend up to 100~pc, while the inner radius of tori is probably of 
the order of half a light year for a 10$^{38}$~Watts AGN \citep{Kitchin2007}. 
We thus use an inner radius of 0.25~pc and an outer radius of 100~pc for 
our first set of torus models. In Paper~II, we concluded that 
``a wide half-opening angle ($\sim$~60$^\circ$) for the torus helps to 
produce parallel polarization, whereas narrow tori [...] produce 
polar-scattering-dominated AGNs''. In this paper, we still model 
circumnuclear dust distributions with half-opening angles 
30$^\circ~\le~\theta_0\le$~60$^\circ$ to check if clumping can have a 
positive impact on our previous conclusions about uniform-density tori. 
Similarly to Paper~II, our models have an optical depth in the V-band of 
$\tau_{\rm dust}$~=~750 along the radial direction. In the case of 
fragmented models, the radius of the spherical gas clouds is fixed to 2~pc 
for all half-opening angles. We vary the number of clouds, from 350 to 
3500, in order to preserve $\cal{F}$. All the clouds, independently of 
the model, have an optical thickness $\tau_{\rm dust}$ = 60, which is in 
agreement with the value ($\ge$~60) used by \citet{Nenkova2008a}.
Our torus models are considered as optically and Compton\footnote{
Around 10 -- 18~$\mu$m, an apparent optical depth of 0.4 corresponds 
to an hydrogen column density of n$_{\rm H}$~=~2$\times$10$^{22}$~cm$^{-2}$ 
\citep{Levenson2014}. A torus is considered as Compton thick when 
n$_{\rm H} \ge$~10$^{24~}$cm$^{-2}$.} thick \citep{Kartje1995}. 
Finally, the clumps and the uniform-density tori are filled with 
dust grains based upon the Milky Way mixture \citep{Draine1984}.

   \begin{figure} 
   \centering
   \includegraphics[width=9.8cm]{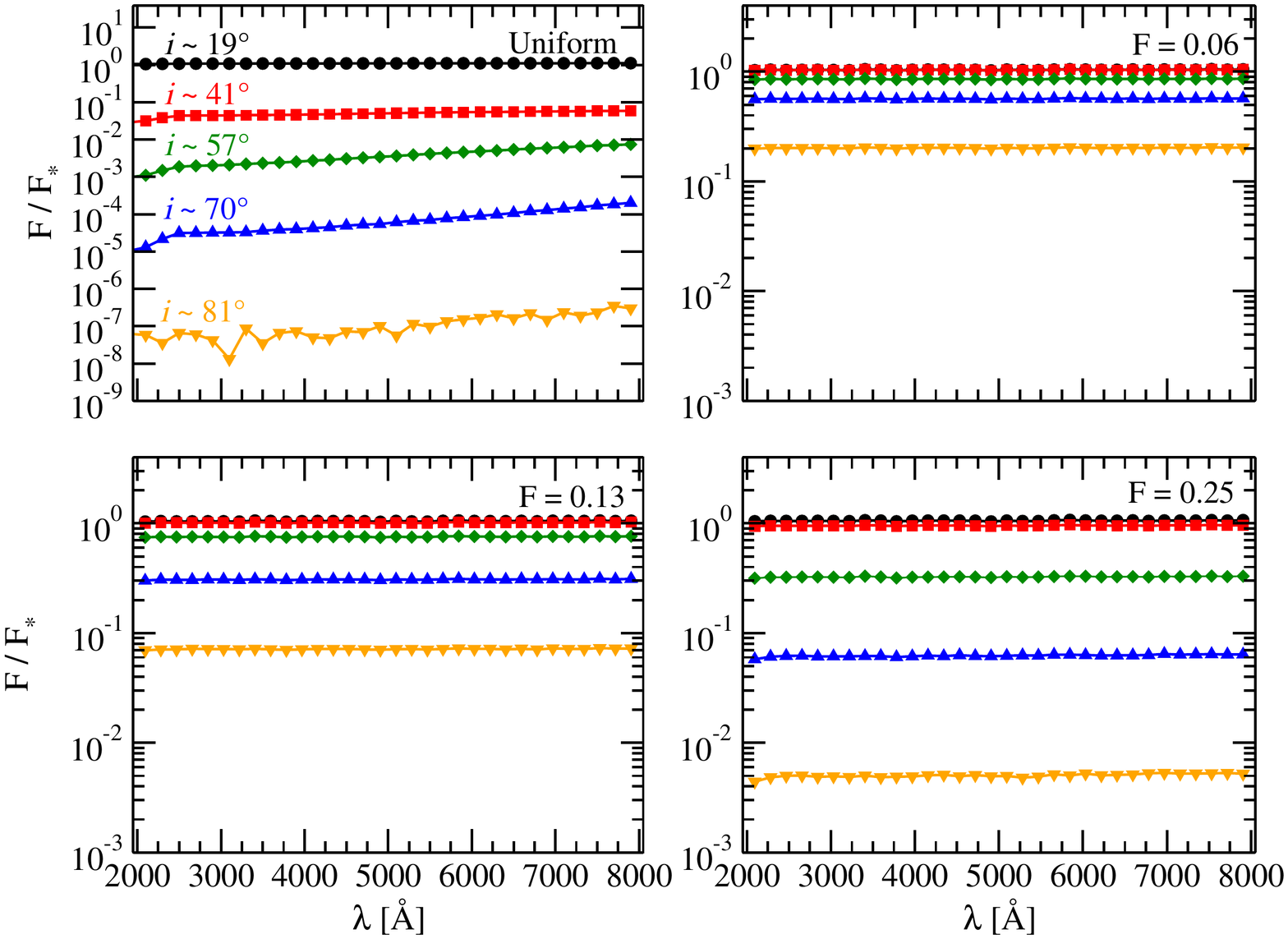}
   \includegraphics[width=9.8cm]{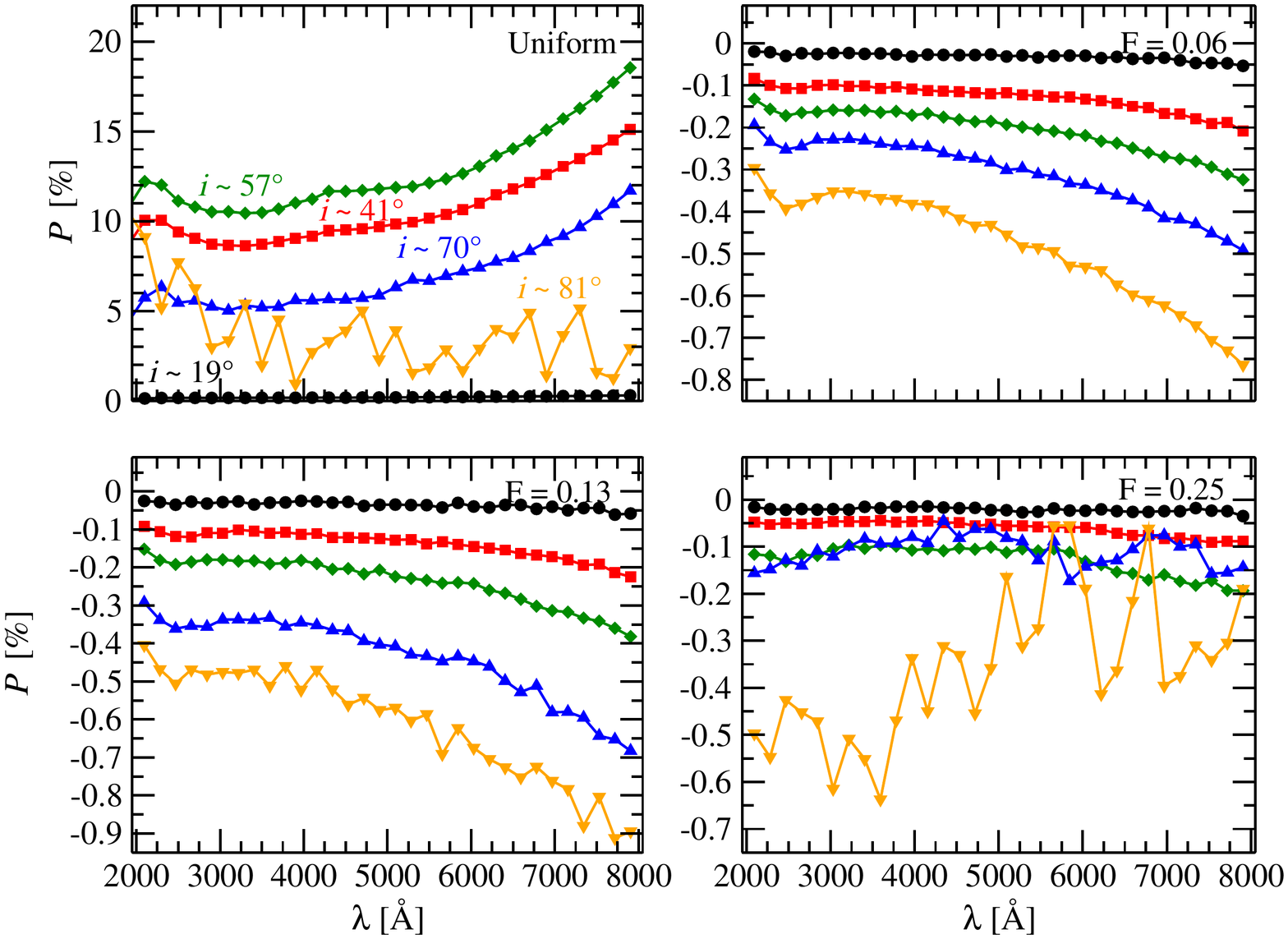}
      \caption{Modeling an optically-thick, elliptically-shaped, extended and fragmented 
		torus with $\theta_0$~=~30$^\circ$ measured relative to the symmetry axis.
		The fraction $F/F_{\rm *}$ of the central flux (upper four panels) and
		the spatially integrated polarization \textit{P} (lower four panels) are seen
		at different viewing inclinations, \textit{i}, from
		2000 to 8000~\AA. The fluxes and polarizations are shown for
		a uniform-density model and three fragmented counterparts
		with filling factors ranging from 0.06 to 0.25.
		The reader is cautioned to notice the 
		change in scale between the spectra of the four models.}
     \label{Fig:Torus30}%
   \end{figure}
%

   \begin{figure} 
   \centering
   \includegraphics[width=9.8cm]{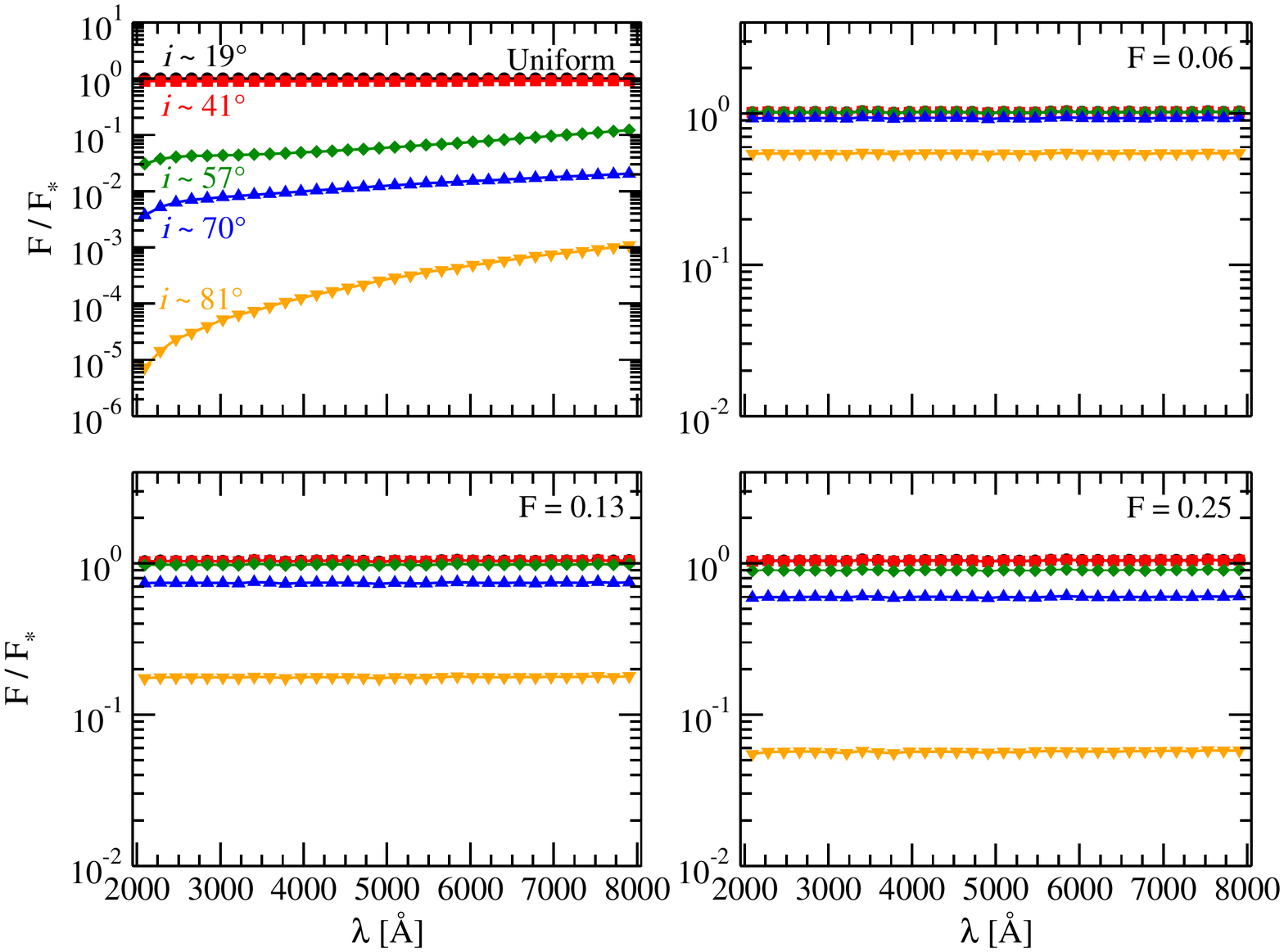}
   \includegraphics[width=9.8cm]{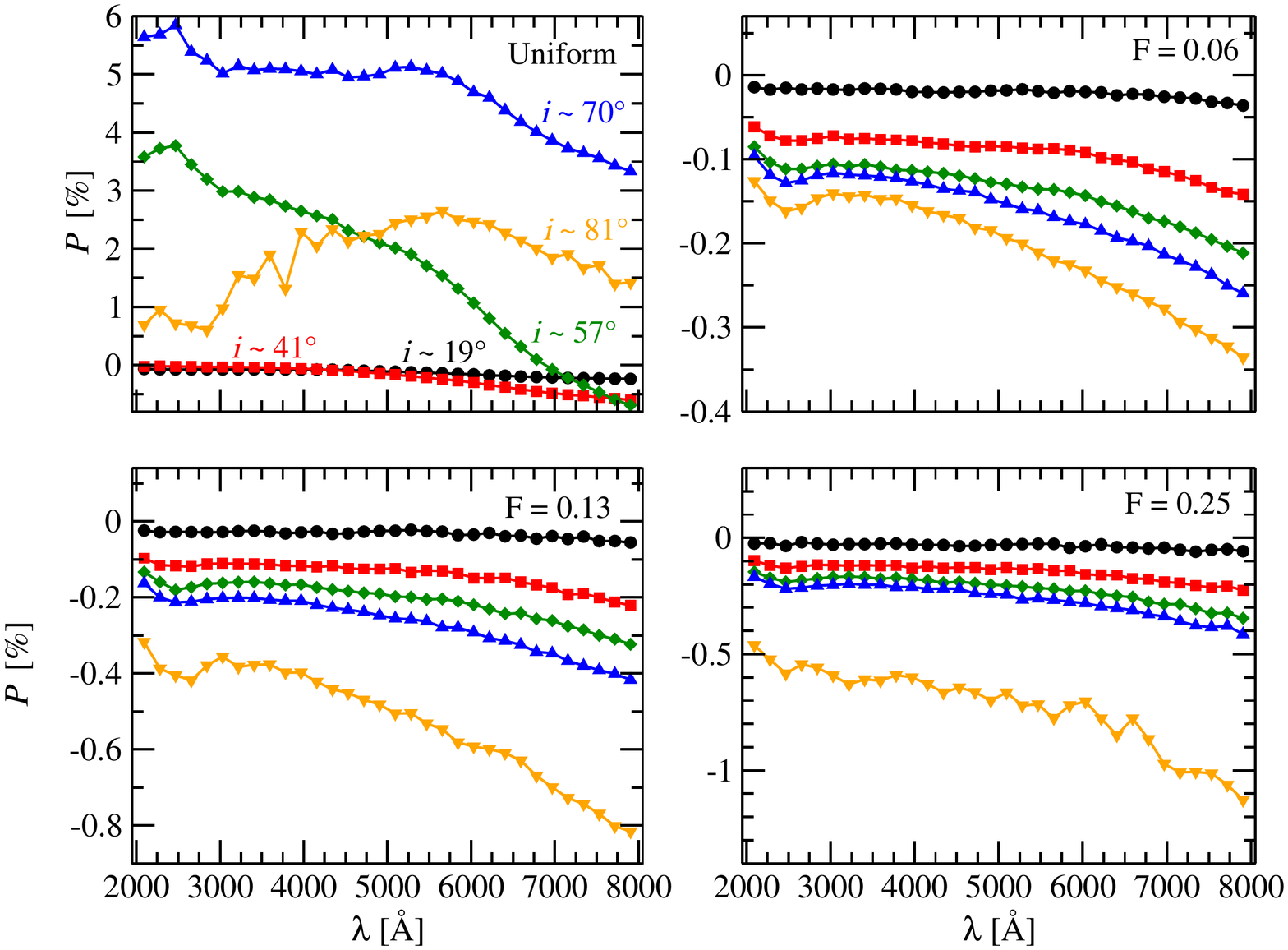}
      \caption{Same as in Fig.~\ref{Fig:Torus30} except that the torus 
		has an opening angle of $\theta_0$~=~45$^\circ$.}
     \label{Fig:Torus45}%
   \end{figure}
%

   \begin{figure} 
   \centering
   \includegraphics[width=9.8cm]{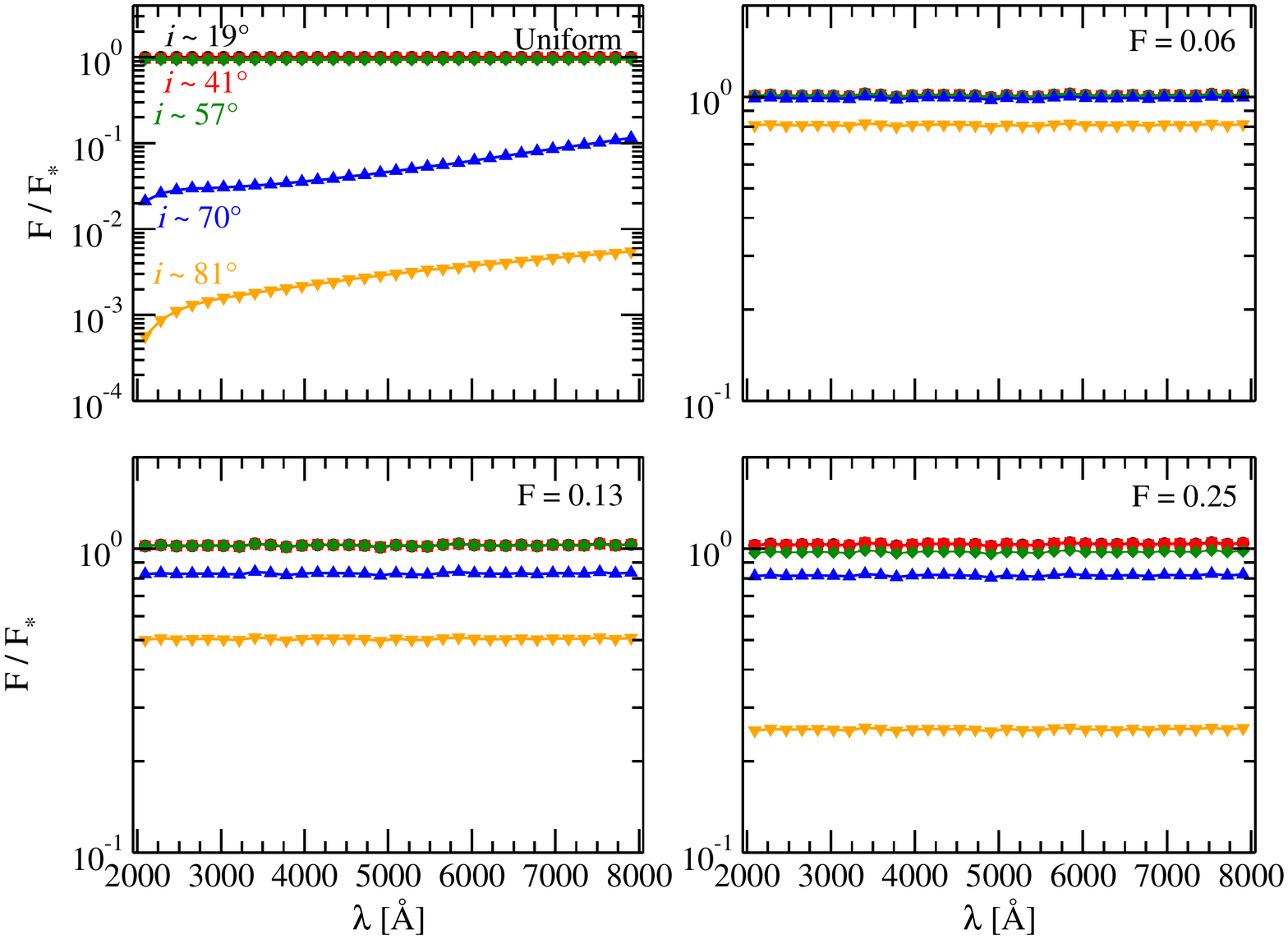}
   \includegraphics[width=9.8cm]{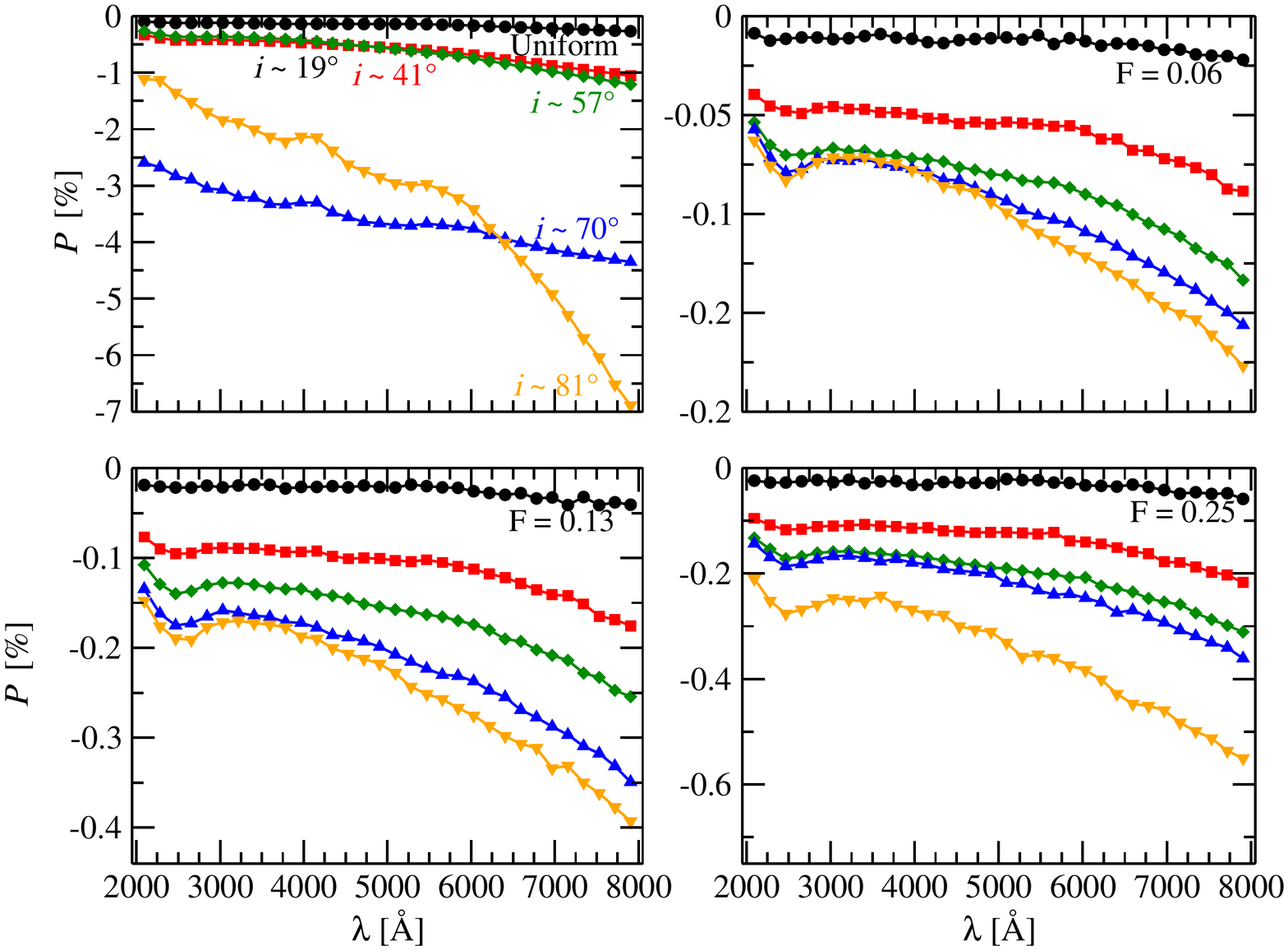}
      \caption{Same as in Fig.~\ref{Fig:Torus30} except that the torus 
		has an opening angle of $\theta_0$~=~60$^\circ$.}
     \label{Fig:Torus60}%
   \end{figure}
%

The fractions $F/F_{\rm *}$ of the central flux, $F_{\rm *}$, are presented in Fig.~\ref{Fig:Torus30},
Fig.~\ref{Fig:Torus45} and Fig.~\ref{Fig:Torus60} (top panels), for $\theta_0$ = 30$^\circ$,
45$^\circ$ and 60$^\circ$ respectively. In comparison with a uniform-density torus, the model with 
$\cal{F} \sim$~0.06 shows, after renormalization, a similar dependence on inclination. The flux is 
maximum for unobstructed viewing angles, i.e. close to the polar directions. With the onset of 
partial obscuration at intermediate inclinations, $F/F_{\rm *}$ shows lower fluxes. 
Maximum attenuation is observed at an equatorial viewing angle ($i \sim$ 81$^\circ$), 
independently of the half-opening angle, but models with $\cal{F} \sim$~0.06 are found to be 
unable to efficiently block photons escaping from the central source. The net type-2 fluxes are 
high in comparison with the flux of the uniform-density model. We also note that the spectral 
shape is less curved (grayer), since radiation can escape in between the clumps, leading to a 
lower absorption efficiency. With higher $\cal{F}$, the partial covering of the source becomes more 
important for intermediate and edge-on views. The dependence of opacity with respect to inclination 
results in lower fluxes at type-2 inclinations. However, even a model with $\cal{F} \sim$~0.25 
cannot completely cancel the escaping flux due to multiple scatterings between the clumps, leading to
a non-null probability of observing the inner AGN regions at extreme orientations \citep{Elitzur2008}.
It indicates that, for a clumpy torus with a medium filling factor, some optical AGN flux 
could be detected at high inclinations, if not diluted by the star light and starburst 
radiation from the host galaxy. If there is contamination by starlight, then the intrinsic 
luminosity of the obscured nuclei has to be renormalized for observations.

The impact on the \textit{polarimetric} signatures of extended clumpy tori, see Fig.~\ref{Fig:Torus30},
Fig.~\ref{Fig:Torus45} and Fig.~\ref{Fig:Torus60} (bottom panels), is greater. Any model with 
$\cal{F} \sim$~0.06 gives polarization about ten times lower than for a uniform-density torus. In 
addition, for $\theta_0$ = 30$^\circ$ and 45$^\circ$, the sign of $P$ changes from positive (perpendicular 
polarization) to negative (parallel polarization) between a uniform-density and a clumpy torus, respectively. 
In the case of $\theta_0$ = 60$^\circ$, the uniform-density torus already induces negative polarization 
due to its scattering geometry, where photons scatter off the side walls of the dust structure and 
naturally yield $\gamma$ values parallel to the torus axis. It is an effect already predicted by 
\citet{Kartje1995} and illustrated here. The spectral shape of polarization is similar to our 
previous results (Papers~I and II): the 2000 -- 3000~\AA~dust feature (specific to the dust 
prescription, see \citealt{Draine1984} and \citealt{Goosmann2007}) and the increase of polarization 
in the red tail of the spectra are properly reproduced by torus models with $\cal{F} \sim$~0.06.
$P$ is minimum at polar viewing angles, where source photons can directly escape from the torus funnel. 
At intermediate torus inclinations, the escaping polarization is higher and is maximum towards the edge 
of the model. There are sufficiently large gaps between the clouds of the fragmented models for the 
radiation to escape along the equatorial plane, contributing to the net polarization. A similar behavior 
is seen for the models with $\cal{F} \sim$~0.13, where $P$ is inferior to 1~\% at type-2 inclinations. 
Finally, when the cloud distributions properly cover the central source ($\cal{F} \ge$ 0.25), 
a large fraction of the photons is absorbed due to dense obscuration and the few escaping photons carry a 
larger $P$ (up to 1~\% for $\theta_0$ = 45$^\circ$). Yet, coupled to a small $F/F_{\rm *}$, the resulting 
polarized flux from equatorial inclinations is expected to be negligible. In the case of a model with 
$\theta_0$ = 30$^\circ$, the dust feature vanishes while it is preserved in the 45$^\circ$ and 60$^\circ$ 
models. The insufficient statistics prevents us to estimate the average polarization degree at a viewing angle 
$i$ = 81$^\circ$ in the former case.

The reprocessing by extended (outer radius = 100~pc), fragmented tori is thus different from the 
case of uniform-density regions with smooth boundaries. While some spectroscopic characteristics such 
as the spectral shape and the graphite peak are conserved (even though the spectra of fragmented tori 
are grayer), polarization can clearly break a degeneracy of the torus reprocessing. The parallel orientation 
of the photon position angle is explained by the angular dependence of opacity and by the small number of 
clouds at the torus maximum height. Due to the oblate geometry of the extended torus, most of the cloudlets 
are concentrated along the equatorial plane and scattering preferentially produces parallel polarization.

\subsubsection{Compact tori}
\label{Clumpy:Impact:Torus_Compact}

In the previous subsection, we explored the polarimetric signatures of
an extended torus. However, the morphology of the dusty material
around AGN is still a matter of debate and several authors have
preferred more compact models: to fit the SED of the quasar 3C249.1,
also known as PG~1100+772, \citet{Heymann2012} used a compact
(external radius: 6~pc) AGN dust torus model composed of more than
5000 optically-thick clouds. \citet{Goosmann2011} also modeled a
compact toroidal structure (external radius: 0.5~pc) to evaluate the
expected X-ray polarization emerging from the Seyfert-2 galaxy
NGC~1068. The difference in the geometry of the extended and compact 
tori can be seen in Fig.~\ref{Fig:Models}. For similar opening angles 
(as shown in Fig.~\ref{Fig:Models}) and similar filling factors, the 
main difference between the compact and extended torus is that the 
inner wall is steeper for the compact torus. 

To investigate the impact of a non-extended torus on the
resulting polarization, we ran simulations for a more
compact model based on the work of \citet{Heymann2012}.
Following their prescription, we compute a torus model with an
inner radius of 0.4~pc and outer radius 6~pc. The half-opening angle
of the torus is set to 45$^\circ$ with respect to the symmetry axis
of the torus and we compare two realizations. The first model is
uniform, with $\tau_{\rm dust}$~=~200 along the radial direction and
the clumpy model is composed of thousands of spheres with R$_{\rm sphere}$ =
0.2~pc, $\cal{F} \sim$~0.25 and optical thickness $\tau_{\rm dust}$ ~=~50. 
The total optical depth along the equatorial plane is thus similar for the 
two cases.

   \begin{figure} 
   \centering
   \includegraphics[width=9.8cm]{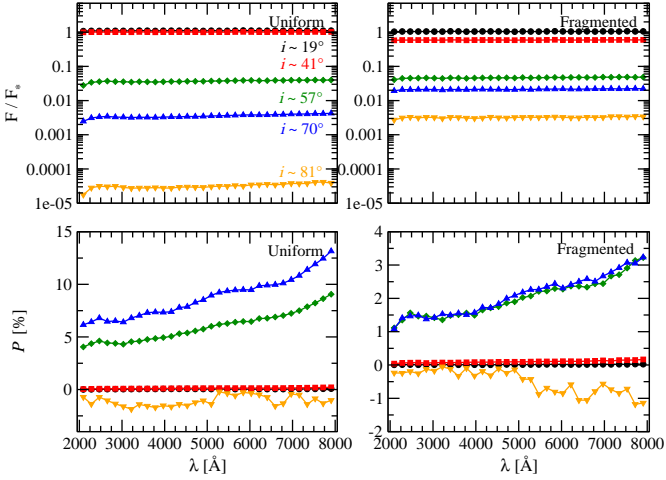}
      \caption{Modeling compact tori (left: uniform-density model; right: clumpy torus) 
		with $\theta_0$~=~45$^\circ$ measured relative to the symmetry axis.
		In the case of the fragmented torus, the filling factor is equal 
		to 0.25	and the optical thickness per clump is 50 in the V-band.
		The fraction $F/F_{\rm *}$ of the central flux (upper figures) and
		the spatially integrated polarization \textit{P} (lower figures) are seen
		at different viewing inclinations, \textit{i}, from
		2000 to 8000~\AA. The reader is cautioned to notice the 
		change in scale between the polarization spectra of the two models.}
     \label{Fig:Torus_compact}%
   \end{figure}
%

Results are given in Fig.~\ref{Fig:Torus_compact}. The spectroscopic
behavior of a toroidal model does not change between a compact
(Fig.~\ref{Fig:Torus_compact} top-left) and an extended
(Fig.~\ref{Fig:Torus45} top panel) medium, regardless of its uniformity;
pole-on inclinations produce the highest fluxes while $F/F_{\rm *}$ is
a minimum at the equator. However, the polarization percentages
between a compact (Fig.~\ref{Fig:Torus_compact} bottom-left) and an
extended uniform-density torus (Fig.~\ref{Fig:Torus45} bottom panel) are
quite different: a compact, uniform, torus produces about twice as much 
polarization with respect to its extended counterpart. This is due to the 
different geometries of their inner walls. The steeper walls of the compact
torus favors the production of perpendicular polarization, except along 
the equatorial direction where scattering of the surface most directly 
facing the observer and producing parallel polarization prevails.
In the case of an extended torus, this surface is more extended as 
the inner torus walls are shallower, resulting in a canceling mix of 
parallel and perpendicular polarization, ultimately lowering $P$. 
However, in agreement with the work of \citet{Kartje1995} and Paper~I, 
both uniform-density models mainly produce a net perpendicular polarization 
angle.

More important changes appear for compact clumpy structures seen through
the prism of polarization. While an extended, fragmented circumnuclear
region only produces parallel polarization with polarization
less than $\sim$ 1~\% (Fig.~\ref{Fig:Torus45}, bottom panel), a compact model 
produces perpendicular polarization together with $P$ up to 3.5~\% 
(Fig.~\ref{Fig:Torus_compact}, bottom right figure). It is a purely 
geometrical effect: the torus being smaller in diameter while sustaining 
a similar height (scaled with the half-opening angle), the torus has a less 
oblate morphology. Hence, the angular dependence of opacity provides higher 
obscuration at intermediate inclinations and radiation does not longer scatter 
preferentially along the midplane, resulting in polarization position angles 
orthogonal to the scattering plane.

Thus, we find that while compact and extended clumpy tori produce similar flux 
levels, their polarization properties are rather different: in the case of an 
extended torus, $P$ is about three times smaller and its polarization angle is 
orthogonal to the one produced by a clumpy compact torus. We thus expect a different 
net polarization signature of AGNs depending on the torus model used. A detailed 
comparison of different fragmented tori and their impact onto polarization will 
be explored in Sect.~\ref{Clumpy:AGN}.

\subsubsection{Polarization imaging}
\label{Clumpy:Impact:Torus_Images}

   \begin{figure} 
   \centering
      \includegraphics[trim = 8mm 5mm 0mm 10mm, clip, width=7.2cm]{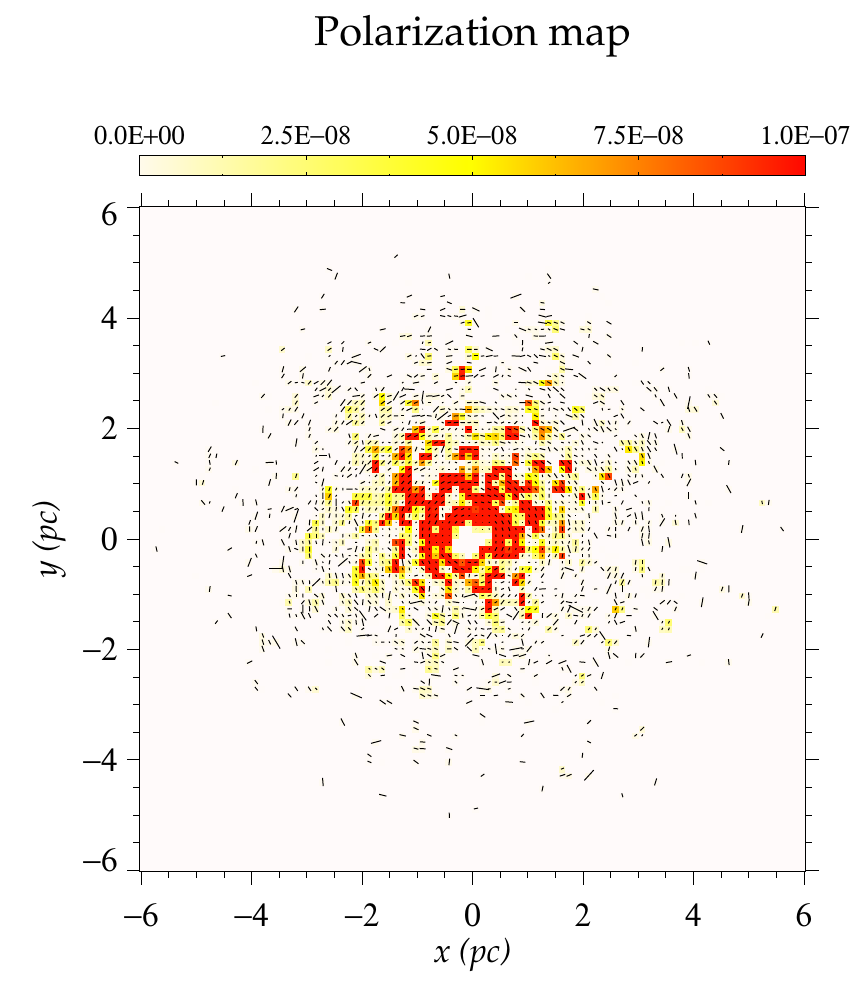}
      \includegraphics[trim = 8mm 5mm 0mm 18.1mm, clip, width=7.2cm]{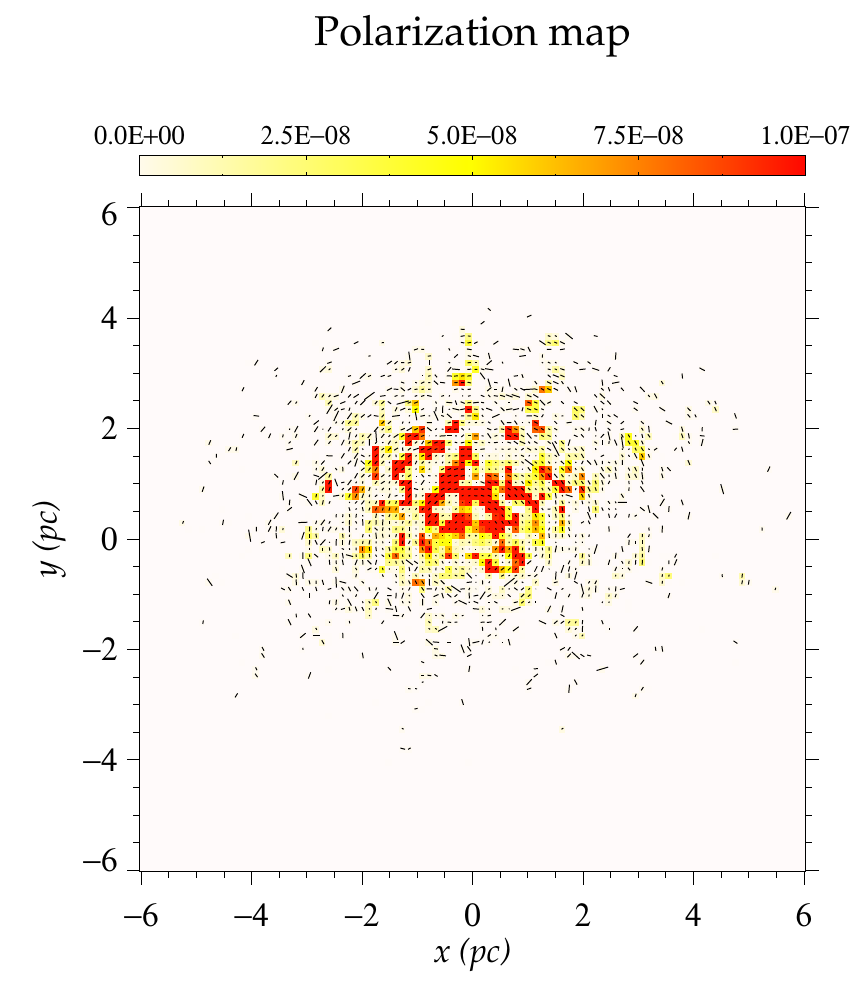}
      \includegraphics[trim = 8mm 5mm 0mm 18.1mm, clip, width=7.2cm]{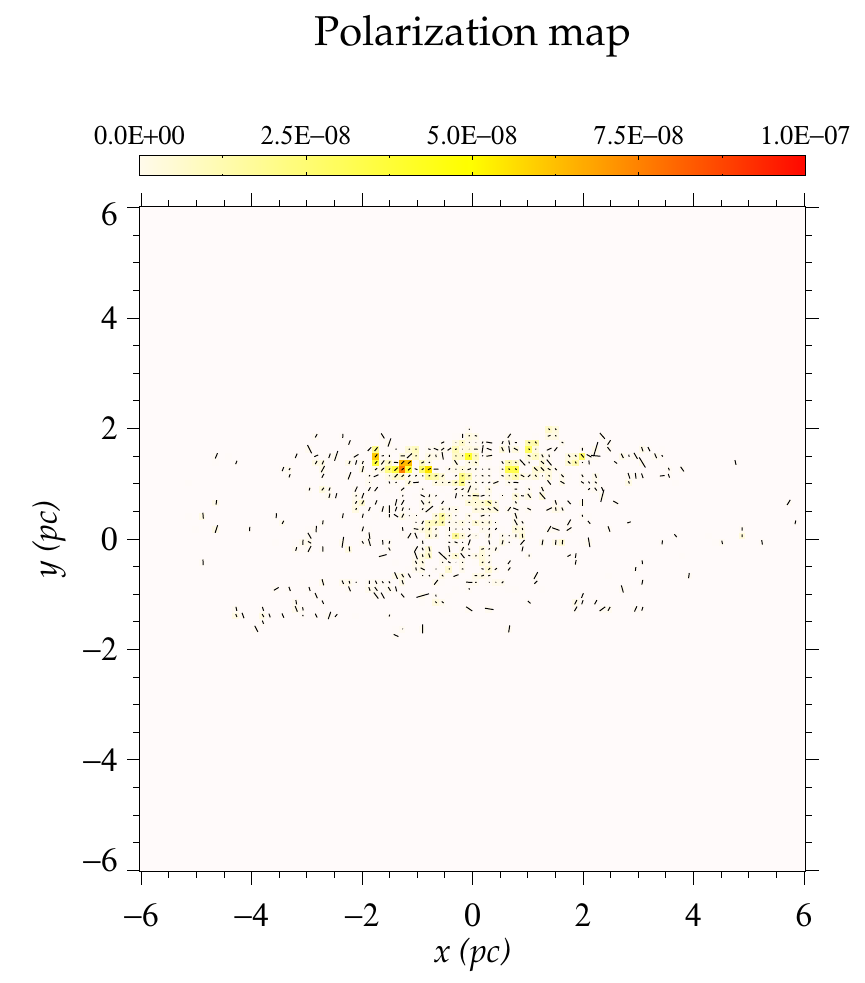}
      \caption{Modeled images of the polarized flux, $PF/F_{\rm *}$, for a
        clumpy, compact dusty torus of half-opening angle 45$^\circ$ measured 
        relative to the axis of symmetry; $PF/F_{\rm *}$ is color-coded and integrated 
        over 2000 to 8000~\AA. $\gamma$ and $P$ are represented by
	black vectors (a vertical vector indicates a polarization of 
	$\gamma = 90^\circ$, and a horizontal vector stands 
	for an angle of $\gamma = 0^\circ$). The length of the vector 
	is proportional to $P$.
        \textit{Top}: image at $i \backsim 18^\circ$ (face-on view);
        \textit{middle}: $i \backsim 45^\circ$;
        \textit{bottom}: $i \backsim 81^\circ$ (edge-on view).}
     \label{Fig:Torus_Map}%
   \end{figure}
%

Using the imaging capabilities of the latest version of {\sc stokes}
(Paper~II), we compute the polarization maps of a compact torus model
such as presented in Sect.~\ref{Clumpy:Impact:Torus_Compact}. We
sample 10$^9$ photons distributed in 10$^4$ spatial bins (100
$\times$ 100 pixels). Each pixel stores the spectra of the four Stokes
parameters across a wavelength range of 2000 to 8000~\AA. Integrating
over this range, the images simultaneously show $PF/F_{\rm *}$, $P$,
and $\gamma$. The polarization position angle $\gamma$ is represented
by black vectors drawn in the center of each pixel; the length of the
vector being proportional to $P$. The black vectors rotate from a
vertical (parallel polarization) to a horizontal position
(perpendicular polarization).

Fig.~\ref{Fig:Torus_Map} presents imaging results at three different
inclinations: 18$^\circ$, 45$^\circ$ and 81$^\circ$, representative of
a type-1 view, a line of sight grazing the torus height, and a type-2
inclination respectively. The peculiar pole-on and edge-on 
values we choose reflect the geometric division of our model space, 
as our results are recorded as a function of $\cos{i}$, where $i$ is 
measured from the axis of the torus (see Paper~I). The pole-on view of 
our model (Fig.~\ref{Fig:Torus_Map} top figure) allows us to probe the 
inner funnel of the circumnuclear matter, where most of the polarized flux
is concentrated. Due to the clumpy structure of the model, the torus 
funnel has no smooth, continuous walls; radiation sweeps
between the cloudlet distribution and reaches farther portions of the
medium. The polarization map clearly shows individual polarized flux
knots at radial distances up to 2.5~pc from the full continuum. In
comparison with a uniform-density torus model (see Fig.~4 in Paper~II) where
most of the polarized photons are absorbed at distances inferior to
0.5~pc, a clumpy structure allows the radiation to penetrate farther
into the inner structure of the torus. The photon flux decreases at
large distances from the central engine, producing polarized signatures
up to 5~pc. In Fig.~\ref{Fig:Torus_Map} (middle), the inner funnel
disappears behind the torus horizon, even if the structure is
fragmented. The optical thickness of the clouds prevents a direct view
of the inner parts of the AGN, but allows photons to escape at large
distances from the source after cumulative scattering events
(increasing the resulting $P$). Along the equatorial plane
(Fig.~\ref{Fig:Torus_Map} bottom), $PF/F_{\rm *}$ sharply decreases
as it is heavily obscured by dust particles. The residual fluxes emerge from
multiple scattering between the gas clumps and the slightly more
polarized flux appearing at the top of the structure is due to the non
orthogonal inclination of the line of sight (81$^\circ$), resulting
in an asymmetric pattern between the top and bottom parts of the
torus. Several clouds can be detected due to higher $PF/F_{\rm *}$
knots but most of the structure is opaque.

\subsection{Fragmented polar outflows}
\label{Clumpy:Impact:Winds}

Investigating the clumpiness of AGN outflows may lead to the most
relevant observational predictions that we can realize, as direct
comparisons can be undertaken with past detections from the Hubble
space telescope (HST), when the COSTAR-corrected faint object camera
(FOC) was mounted\footnote{The FOC (1150 -- 6500~\AA) was removed
 from the HST during service mission 3B in March, 2002}. We now
look at two different realizations of the winds, the first one using
electron-filled spheres to mimic the ionized winds and the second one
using dust clouds to represent the more extended NLR.

\subsubsection{Ionization cones}
\label{Clumpy:Impact:Winds:Electron}

Following the numerous observations of AGN ionization cones 
presented in \citet{Wilson1996}, we model a 10~pc-long, hourglass-shaped 
scattering region filled with electron spheres of radius 
R$_{\rm sphere}$~=~0.27~pc and optical depth $\tau_{\rm elec}$~=~0.3 
(in comparison, the uniform-density model has $\tau_{\rm elec}$~=~1). 
Since the cloudlets are filled with absorption-free, 
optically thin material, the effective optical thickness along the radial 
direction is expected to vary between the different filling factors. 
We evaluate $\tau_{\rm total}$ to be roughly similar between the uniform-density 
and the models with $\cal{F} \sim$~0.06, while $\tau_{\rm total}$ is closer 
to unity for the two models with higher filling factors. The half-opening 
angle of the double cone is again set to $30^\circ$. The emitting source is 
not restricted to radiate along a preferred direction aligned with the cones 
as for the model described in Paper~I but emits isotropically.

   \begin{figure} 
   \centering
   \includegraphics[width=10.5cm]{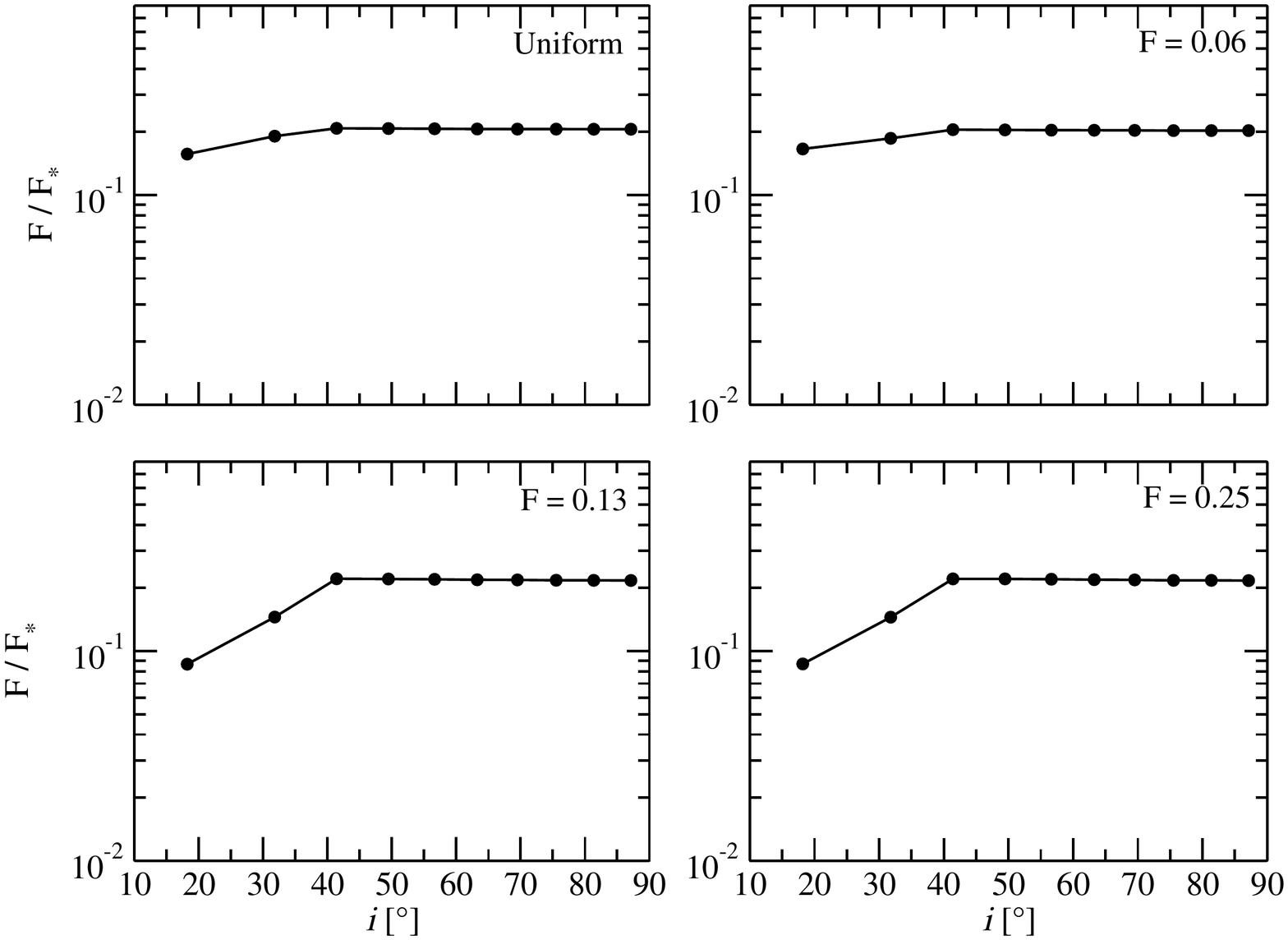}
   \includegraphics[width=9.9cm]{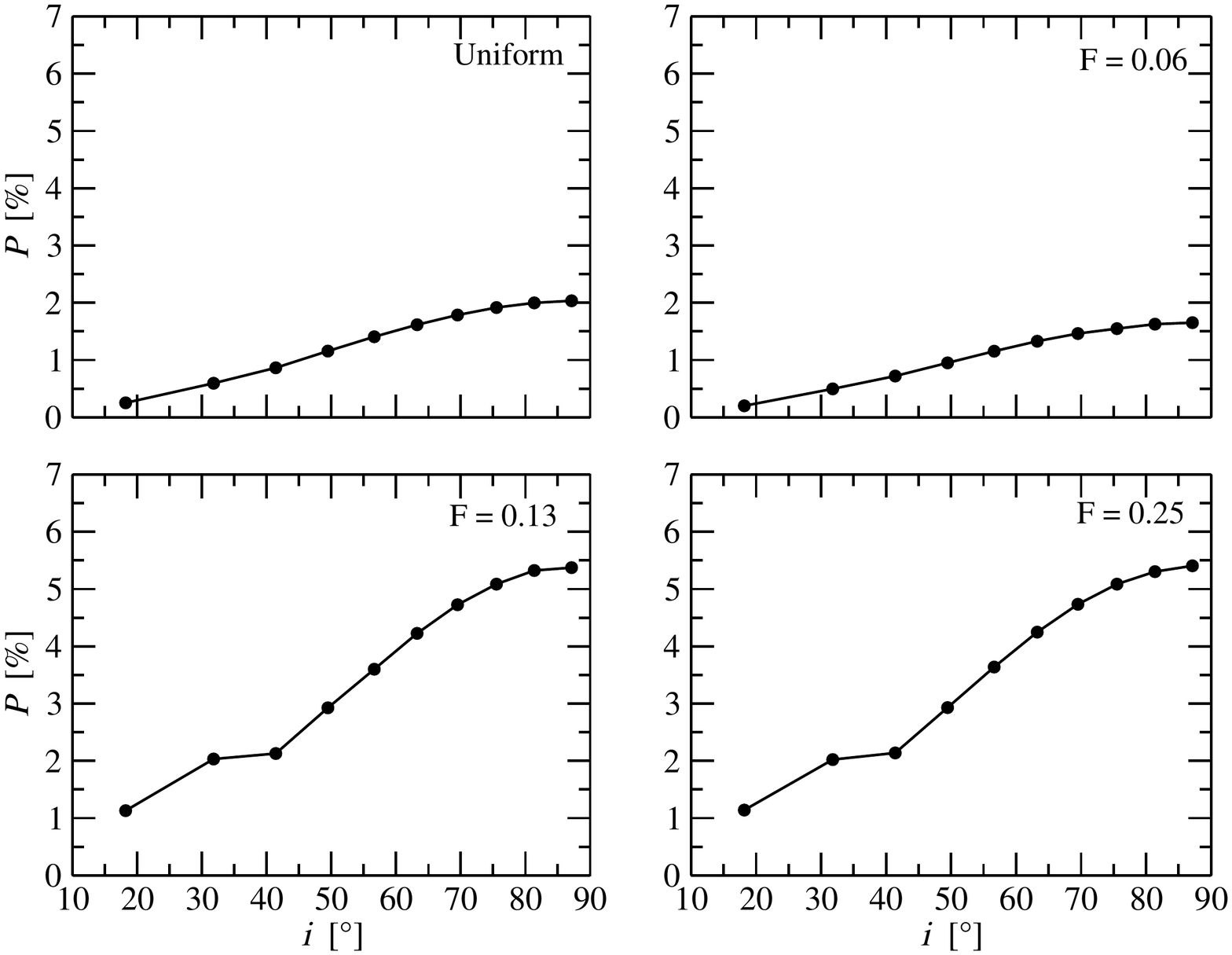}
      \caption{Modeling an electron-filled, fragmented, scattering double-cone
		with $\theta_0$~=~30$^\circ$ measured relative to the symmetry axis.
		The fraction $F/F_{\rm *}$ of the central flux is integrated from
		2000 to 8000~\AA~(upper panel) and the spatially integrated polarization 
                \textit{P} is plotted versus inclination \textit{i} 
		(lower panel).}
     \label{Fig:Wind_elec}%
   \end{figure}
%

We find that the fraction $F/F_{\rm *}$ of the central flux
(Fig.~\ref{Fig:Wind_elec}, upper panel) is very similar between a
uniform-density and a fragmented model using a low filling factor. 
Due to obscuration of the nuclei by the cloudlet distribution and 
subsequent scattering off the observer's line of sight, less flux
is detected at type-1 inclinations. However, due to the Thomson,
elastic scattering mechanism, the flux difference between type-1 and 
type-2 viewing angles is not important as no absorption occurs. When the
filling factor increases, multiple scattering is enhanced even
when considering optically thin scattering clouds. A larger fraction
of the flux is deviated from the polar direction and $F/F_{\rm *}$
decreases. 

The polarization degree curves (Fig.~\ref{Fig:Wind_elec}, bottom
panel) fit in with this consideration: a model with $\cal{F} \sim$~0.06 
reproduces the integrated polarization degree that we found for a uniform-density 
model, with a maximum $P$ at equatorial viewing angles. The shape of
$P$ with respect to the system's inclination is quite similar too. For
$\cal{F} >$ 0.06, as there are more scattering targets and a higher effective 
optical depth, the resulting polarization percentage increases up to a 
factor three. The inclination dependence is preserved but $P$ increases. 
The polarization position angle is positive for all the
models as the spheres are situated mainly along the vertical axis of the 
model. 

A model of fragmented winds is able to reproduce the same amount of
positive polarization degree, as well as the same $P$-dependency
versus inclination, as a uniform-density model if the filling factor of the
wind is small ($<$~0.1). This threshold is consistent with the radio 
constraints on the volume filling factors of AGN winds found by 
\citet{Blustin2009}. Denser cloud distributions produce higher polarization
degrees due to the multiplication of scattering targets that ultimately increases 
the total optical depth. Uniform models are then good approximations for
real systems containing only a few gas clouds. It is noteworthy 
that this result is in contrast to flux-only measurements, where models 
with higher $\cal{F}$ resemble uniform-density models \citep{Ashton2005}.

\subsubsection{NLR winds}
\label{Clumpy:Impact:Winds:Dust}

We now consider a fragmented outflow consisting of dusty spheres with
R$_{\rm sphere}$ = 2.7~pc and $\tau_{\rm dust}$~=~0.1 (previous
uniform-density model: $\tau_{\rm dust}$~=~0.3). The clouds are optically 
thin in order to represent the NLR clouds detected farther away from 
the irradiating source, where dust can form and survive \citep{Hoenig2012}. 
We consider that the dusty double-cone sustains the same half-opening angle as the 
ionized outflows, $\theta_0$ = $30^\circ$. According to the observations 
of \citet{Hoenig2012} and our simulations in \citet{Marin2012b}, the inner 
and outer boundaries of the reprocessing region were fixed to 10~pc and 
100~pc above the source, respectively.

   \begin{figure} 
   \centering
   \includegraphics[width=9.8cm]{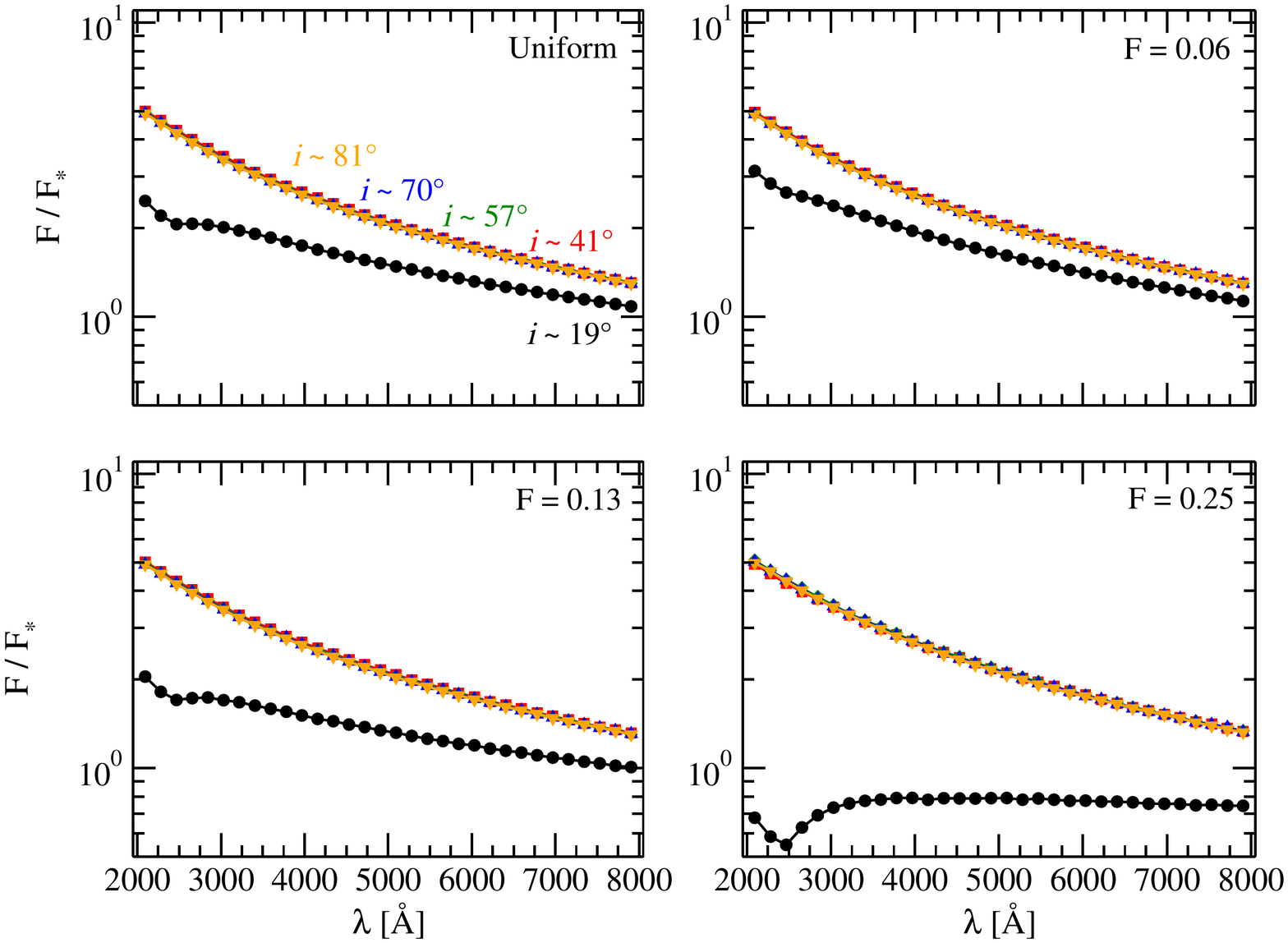}
   \includegraphics[width=9.8cm]{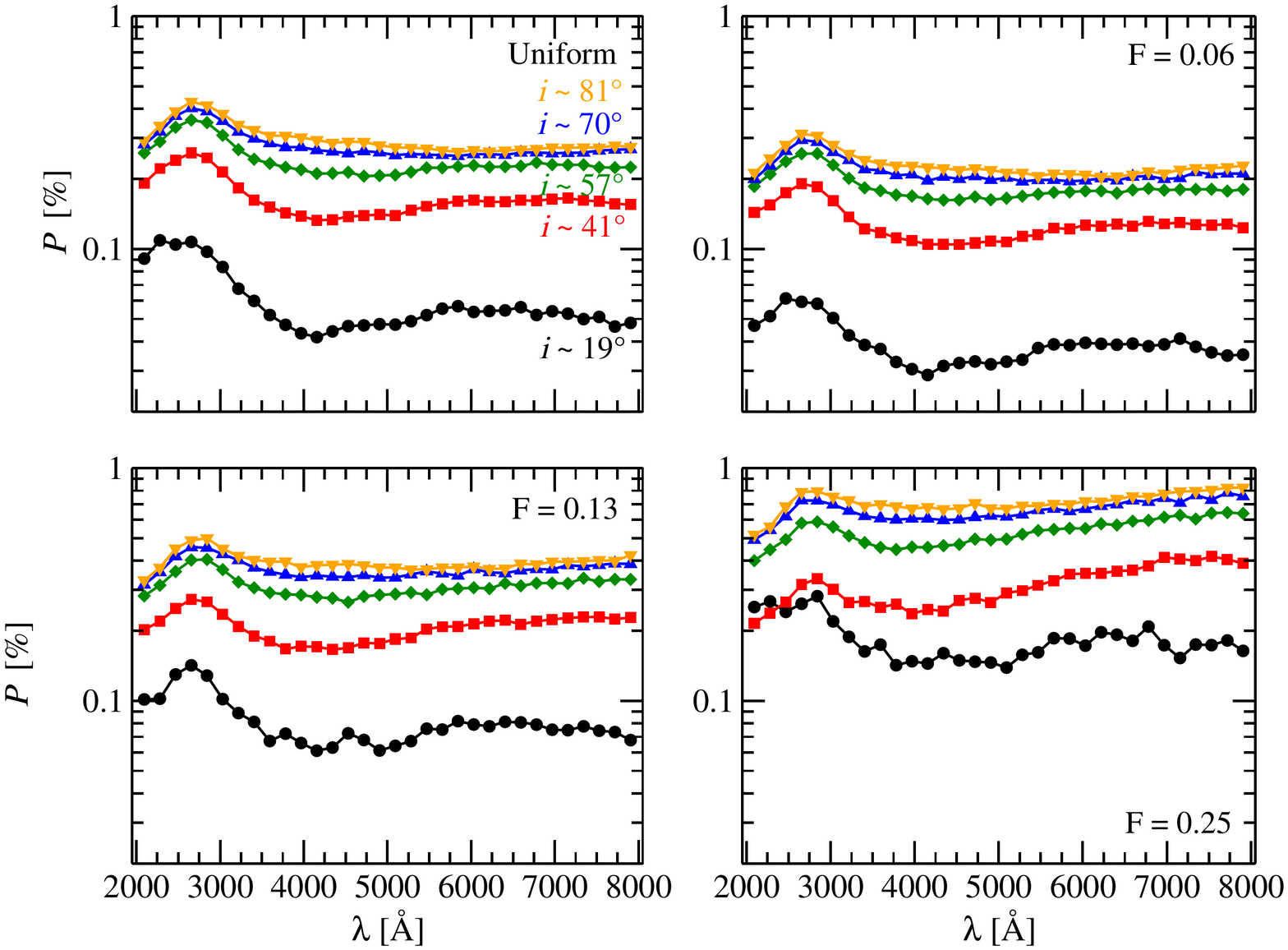}
      \caption{Modeling an optically thin, dusty, fragmented double-cone
		with $\theta_0$~=~30$^\circ$ measured relative to the symmetry axis.
		The fraction $F/F_{\rm *}$ of the central flux (upper panel) and
		the spatially integrated polarization \textit{P} (lower panel) are seen
		at different viewing inclinations, \textit{i}, from
		2000 to 8000~\AA.}
     \label{Fig:Wind_dust}%
   \end{figure}
%

The $F/F_{\rm *}$ spectra produced by scattering on a small number of
spheres (Fig.~\ref{Fig:Wind_dust} upper panel) reproduces the expected
flux behavior: there is an unabsorbed photon flux along lines-of-sight
that do not cross the NLR, and the flux is slightly absorbed at
pole-on directions. The graphite peak signature is marginally
detectable when considering a $\cal{F} \sim$~0.06 model but when looking 
at a denser cloud distribution ($\cal{F} \sim$~0.13 and 0.25), 
the 2000 -- 3000~\AA~dust feature becomes more apparent and the flux of 
radiation escaping from the model at $i \sim$ 19$^\circ$ is lower. The 
$\cal{F} \sim$~0.13 case is found to be the closest one to the 
uniform-density model.

The polarization percentage induced by a fragmented wind with a low
number of scattering targets is in the same range as for a uniform
model (Fig.~\ref{Fig:Wind_dust} lower panel). According to the
scattering phase function of polarization, a maximum polarization
degree is observed for equatorial viewing angles, with the
2000 -- 3000~\AA~peak also revealed in polarization percentage. For 
observer's lines of sight close to the polar direction, $P$
is minimum as absorption by dust grains limits the transmission of
photons through the gas. For a $\cal{F} \sim$~0.06 model, $P$ is
slightly lower than for the uniform-density case, but for a larger number of 
gas clouds, the models are very similar, as expected from Fig.~\ref{Fig:Wind_dust}
(upper panel). For a $\cal{F} \sim$~0.25 model, $P$ is higher than for
a uniform-density regions due to a larger number of scattering targets
and an increased probability for scattered radiation to escape from
the model in between the cloudlets.

It is interesting to note that a polar distribution of spheres does
not strongly impact the resulting polarization degree in comparison
with a uniform-density model. Taking into account a low-to-moderate
number of clumps, one can use a continuous outflow model as a good
approximation for polarimetric modeling. This result does not apply to
equatorial distributions of gas (see Sec.~\ref{Clumpy:Impact:Torus}
and the following section).

\subsection{Disrupted accretion flow}
\label{Clumpy:Impact:Flared}

   \begin{figure}
   \centering
   \includegraphics[width=10.5cm]{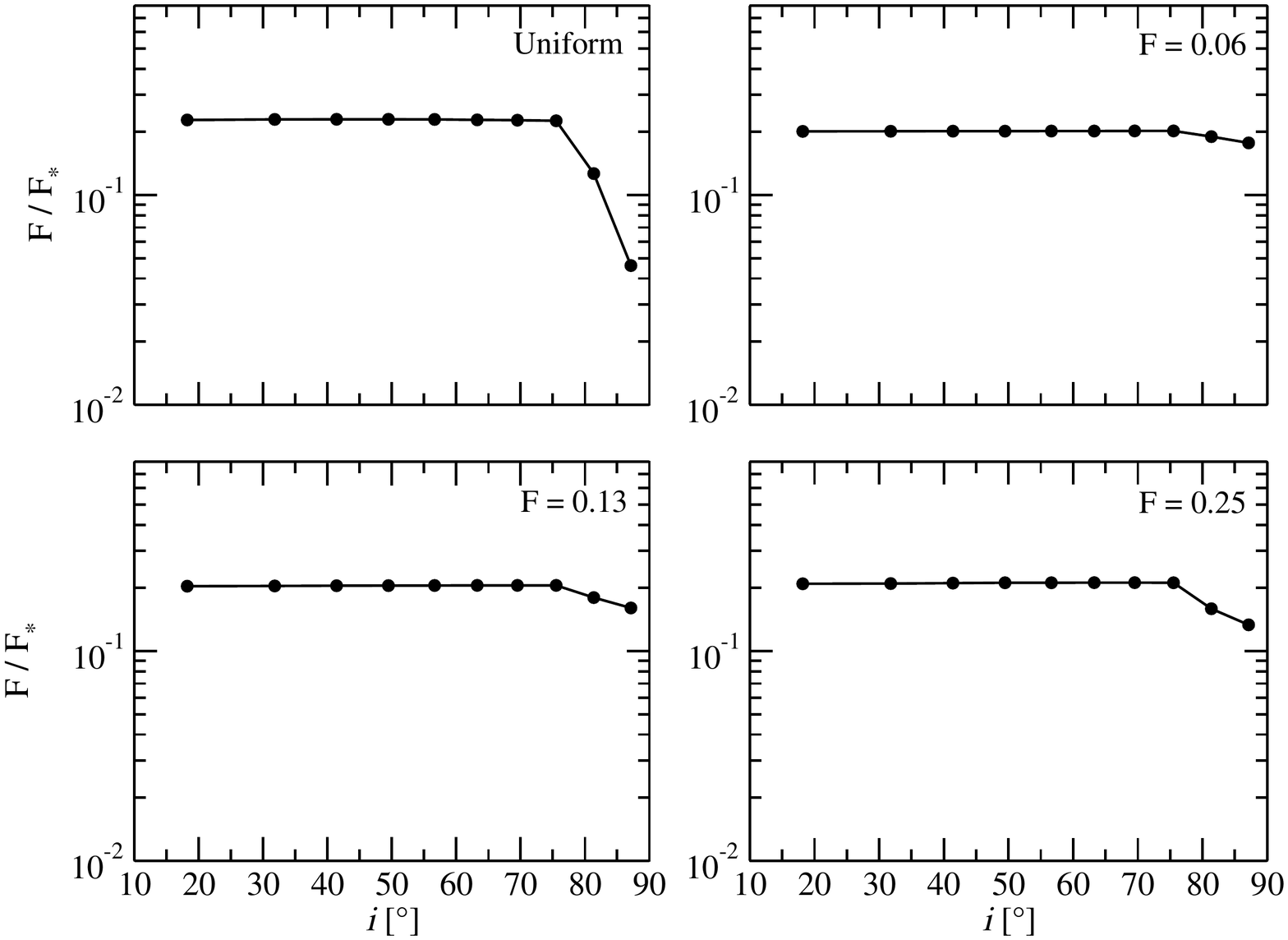}
   \includegraphics[width=10.5cm]{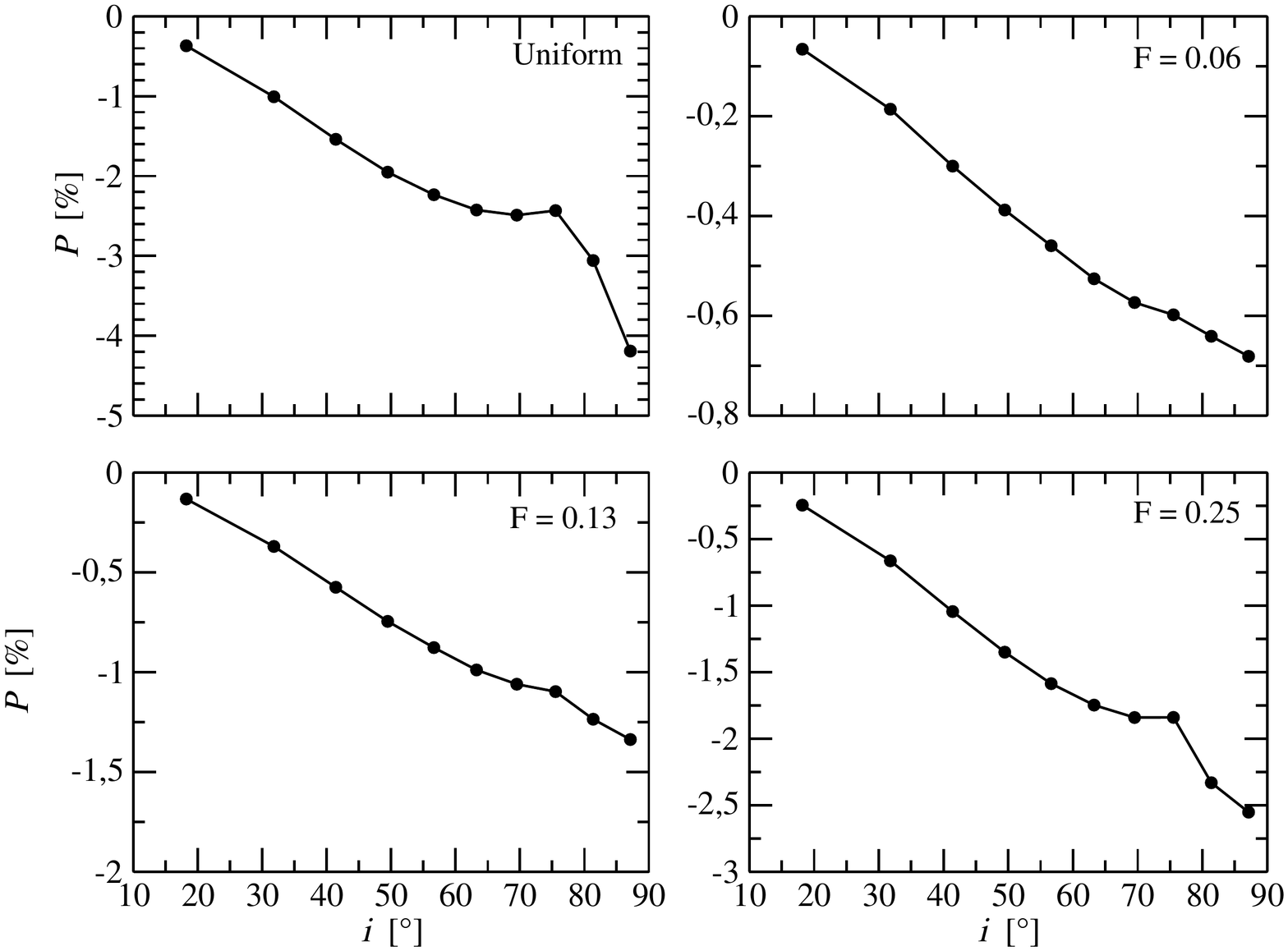}
      \caption{Modeling of an equatorial, electron-filled, fragmented scattering disk with a
		half-opening angle $\theta_0$~=~$80^\circ$.
		The fraction $F/F_{\rm *}$ of the central flux (upper panel) and
		the spatially integrated polarization \textit{P} (lower panel) are seen
		at different viewing inclinations, \textit{i}, integrated over
		2000 to 8000~\AA. The reader is cautioned to notice the 
		change in scale between the polarizations of the four models.}
     \label{Fig:Flared}%
   \end{figure}
%

The last scattering region to be examined is the accretion flow between the 
torus and the accretion disk. According to the simulations of 
\citet{Young2000}, \citet{Smith2004}, and Paper~II, this flow is likely to be a 
flattened region (half-opening angle of $80^\circ$) with a distribution of 
optically thin matter between 3$ \times 10^{-4}$~pc and 5$ \times 10^{-4}$~pc.
We thus model a fragmented equatorial scattering region composed of electron 
spheres with R$_{\rm sphere}$~=~1.17~$\times 10^{-5}$~pc and $\tau_{\rm elec}$~=~0.3 
(in the previous uniform-density model, $\tau_{\rm elec}$ was set to unity). 

Equatorial scattering by a distribution of ionized clouds only impacts
equatorial lines-of-sight (Fig.~\ref{Fig:Flared} upper panel). When
radiation passes through the model, it escapes more easily from a
fragmented model with a low number of scattering targets than from a
uniform-density reprocessing region. Thus, the observed flux is higher along
the equator for sparsely populated models. For larger $\cal{F}$, 
the angular dependence of opacity changes, the partial covering of the 
source is larger, so the flux recorded at $i >$ 75$^\circ$ is minimized. 
But, due to a higher total optical depth and the gaps between clouds, the 
amount of photons reprocessed along the equator is still higher than for 
a uniform model, even when considering a large filling factor.

As photons more easily escape from a fragmented model, the fraction of
reprocessed radiation is lower, so the polarization degree
(Fig.~\ref{Fig:Flared} lower panel) decreases. The general tendency is
the same in comparison with a continuous accretion flow: the
maximum polarization degree is detected at equatorial inclinations, but
its percentage is much lower. A fragmented medium also reproduces the
parallel orientation of the photon polarization angle, similarly to an
extended clumpy dusty torus model. To maximize $P$, equatorial
covering must be higher, and only with $\cal{F} >$ 0.2 the model
can approximatively reproduce the polarization signature of the
uniform-density case.


\section{AGN modeling with clumpy structures}
\label{Clumpy:AGN}

\subsection{Comparison with a uniform-density model}
\label{Clumpy:AGN:uniform}

Now that we have investigated the isolated reprocessing components separately, our last step is to construct a three-component model to approach a typical 
unified scheme for AGNs \citep{Antonucci1993}. To do so, we gather three fragmented regions around the central emitting source. First, the equatorial 
electron-dominated flow is modeled using a $\cal{F} \sim$~0.25 model in order to maximize the production of parallel polarization at polar viewing angles 
\citep{Smith2004}, see Sect.~\ref{Clumpy:Impact:Flared}. Then, at greater distances, a fragmented torus is set to obscure the flux coming from the inner 
part of the AGN. We opted for another $\cal{F} \sim$~0.25 model to enhance the partial covering factor of the source \citep{Honig2010}. However, as noted 
in Sect.~\ref{Clumpy:Impact:Torus}, an extended or a compact, clumpy, circumnuclear medium results in very different polarization signatures. We thus 
opted for two realizations of our AGN model, one with a compact dusty structure and a second with a large torus recovered from our simulations in 
Sect.~\ref{Clumpy:Impact:Torus_Extended} and Sect.~\ref{Clumpy:Impact:Torus_Compact}, respectively. We now set the fragmented tori to the same 
half-opening angles: $\theta_0$~=~30$^\circ$. Finally, the tori are set to collimate the ionized outflows. Following observational evidences for 
polar outflows with low filling factors (e.g. NGC~3516, MR~2251-178 and MCG-6-30-15, \citealt{Blustin2009}) we opt for a wind model with $\cal{F} \sim$~0.06, 
see Sect.~\ref{Clumpy:Impact:Winds:Electron}.

\begin{figure*}
   \centering
   \includegraphics[width=14cm]{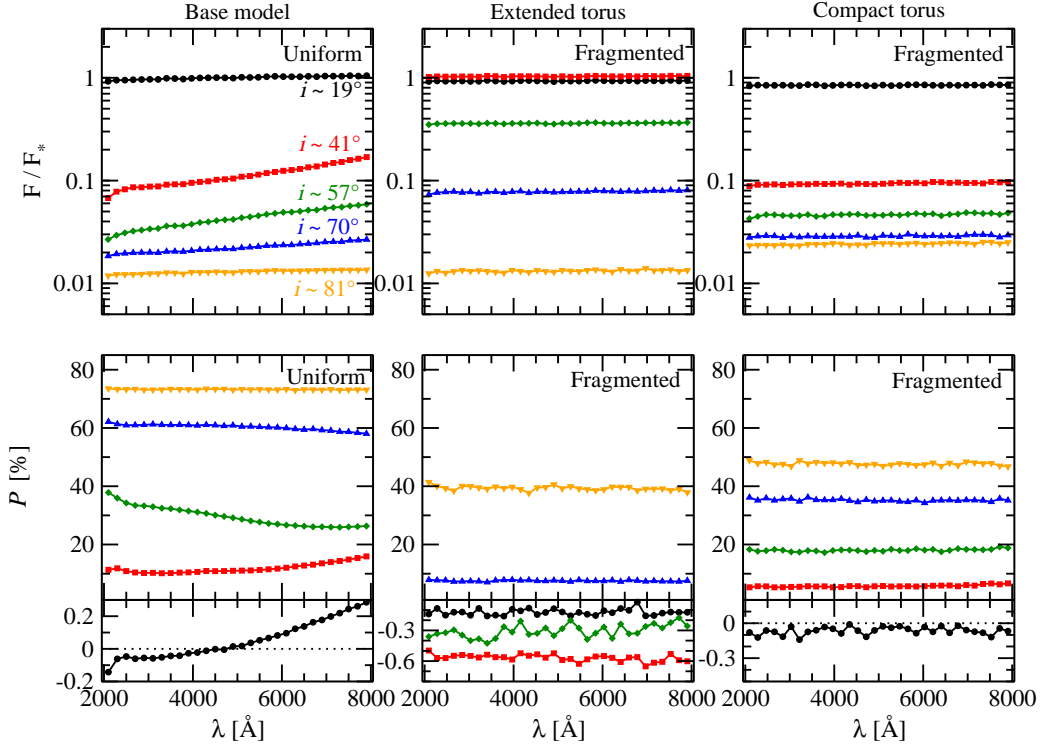}
      \caption{Modeling the AGN unified scheme by three reprocessing regions (see text).
		The fraction $F/F_{\rm *}$ of the central flux is plotted from
		2000 to 8000 \AA~(upper panel) and the spatially integrated polarization 
                \textit{P} is plotted versus the observer's inclination \textit{i} 
		(lower panel). The left column shows an AGN model composed
		of only uniform-density reprocessing regions, while the middle and right columns show
		different AGNs (with an extended torus in the middle and a compact torus on 
		the right side sustaining the same, $\theta_0$~=~30$^\circ$, half-opening angle). 
		The clumpy models are composed of the same three reprocessing regions (regarding 
		the smooth-density AGN) which are now fragmented (radiation supported disk: 
		$\cal{F} \sim$~0.25, torus: $\cal{F} \sim$~0.25, and ionized winds: $\cal{F} \sim$~0.06).}
     \label{Fig:AGN}%
   \end{figure*}

We present the resulting AGN modeling in Fig.~\ref{Fig:AGN}, comparing a model made with purely uniform-density reprocessing regions to two clumpy models with
different torus sizes (extended torus: middle column, compact torus: right column). The fraction $F/F_{\rm *}$ of the central flux is found to be different, 
in terms of intensity, between a uniform-density model (Fig.~\ref{Fig:AGN}, top-left) and its fragmented counterparts. The amount of radiation in the polar direction 
is maximum for all models since the polar winds are optically thin. The flux is seen in transmission and has not suffered many scattering events. 
When the viewing angle increases, $F/F_{\rm *}$ slowly decreases for the clumpy AGN model with an extended torus, but not as fast as in the uniform-density case or the 
clumpy AGN with a compact circumnuclear structure. Part of the photons previously trapped by the inner walls of the uniform-density dusty torus are now able to escape, 
thus contributing to the flux. As soon as the line of sight towards the system crosses the torus horizon, the flux decreases sharply, but is still 100 times larger 
than in the uniform-density case due to the non-perfect obscuration of the nuclei by the gas clumps. Along the equatorial plane, where a minimum amount of radiation 
is detected, photons escape from the model by (mostly) orthogonal scattering in the winds, and by direct transmission through the circumnuclear matter. The resulting 
type-2 $F/F_{\rm *}$ is then much higher for a fragmented model with a large toroidal structure than for the other two models. $F/F_{\rm *}$ is found to be roughly 
similar between the uniform-density and the clumpy AGN model with a compact torus. This is in agreement with our conclusions found in Sect.~\ref{Clumpy:Impact:Torus_Compact}, 
indicating that a compact dusty structure acts in a similar manner as a uniform-density torus, but producing grayer intensity fluxes due to clumping.

%

Fragmentation strongly influences the resulting polarization signatures of an AGN model. In the uniform-density case (Fig.~\ref{Fig:AGN}, bottom-left), 
parallel (negative) polarization is only detected at viewing angles close to the pole ($i~\sim$~19$^\circ$), but in the case of a clumpy model with an extended torus
(Fig.~\ref{Fig:AGN}, bottom-middle) parallel polarization is detected up to $i~\sim$~57$^\circ$. Such an inclination coincides with a viewing angle passing directly 
through the torus clumps without suffering too much absorption. There is a dual contribution from the accretion flow and from the extended dusty torus that are both 
producing negative polarization (see Sect.~\ref{Clumpy:Impact:Torus_Extended}). The levels of $P$ range from 0.1~\% to 0.6~\% for type-1 AGN. When the observer's
line of sight crosses the bulk of the dusty torus, where there is a maximum density of spheres, the photon position angle switches from parallel to 
perpendicular orientation as radiation from the equatorial regions is strongly suppressed. Most of the flux is seen by orthogonal scattering in the ionized winds, 
only producing 0$^\circ$ (perpendicular) polarization angles. $P$ is much lower for $i \sim$ 70$^\circ$ in the fragmented case ($P~\sim$~7~\%), where photons reprocessed 
on the equatorial components are not properly shaded by the torus clumps, than in the case of a uniform-density model ($P~\sim$~60~\%) where polarization originates from
scattering on the winds only. Finally, at an extreme type-2 inclination, the polarization degree is maximum, with $P~\sim$~40~\%. For all the viewing angles, 
$P$ is found to be approximatively wavelength-independent, which is in agreement with the spectropolarimetric observations of NGC~1068 \citep{Antonucci1985}, 
3C~256 \citep{Dey1996}, or 3C~324 \citep{Cimatti1996}. The case of a fragmented model with a compact torus is 
less different in comparison with a uniform-density AGN since it does not favor the production of parallel polarization. Due to the compact scale of the clumpy torus, 
negative polarization is only detected at $i~\le$~30$^\circ$, with a similar degree of $P$. However, the level of perpendicular polarization at intermediate and equatorial 
inclinations is about 1.5 to 2 times lower than for a uniform-density case, since the torus is less effective in producing high degrees of polarization (due to depolarization
induced by multiple scattering on a clumpy distribution of reprocessing regions). At maximum inclinations, $P$ is equal to 50~\%.

Eventually, we find that there are big discrepancies between the two clumpy AGN models, depending on the morphology of their circumnuclear dust region.
For the same half-opening angle ($\theta_0$~=~30$^\circ$), radial opacity ($\tau \gg$~1) and filling factor ($\cal{F}$~=~0.24), an AGN with an extended torus 
produces the smallest $P$ and parallel polarization up to $\sim$~70$^\circ$ while a compact torus mimics the polarimetric behavior of a uniform-density AGN model
but with reduced polarization degrees. We stress that it is due to the geometry of the tori: an extended torus being more oblate than its compact counterpart, thus
favoring the production of parallel polarization.

\subsection{Exploring partially fragmented models}
\label{Clumpy:AGN:Components}

\begin{figure}
   \centering
   \includegraphics[width=9.8cm]{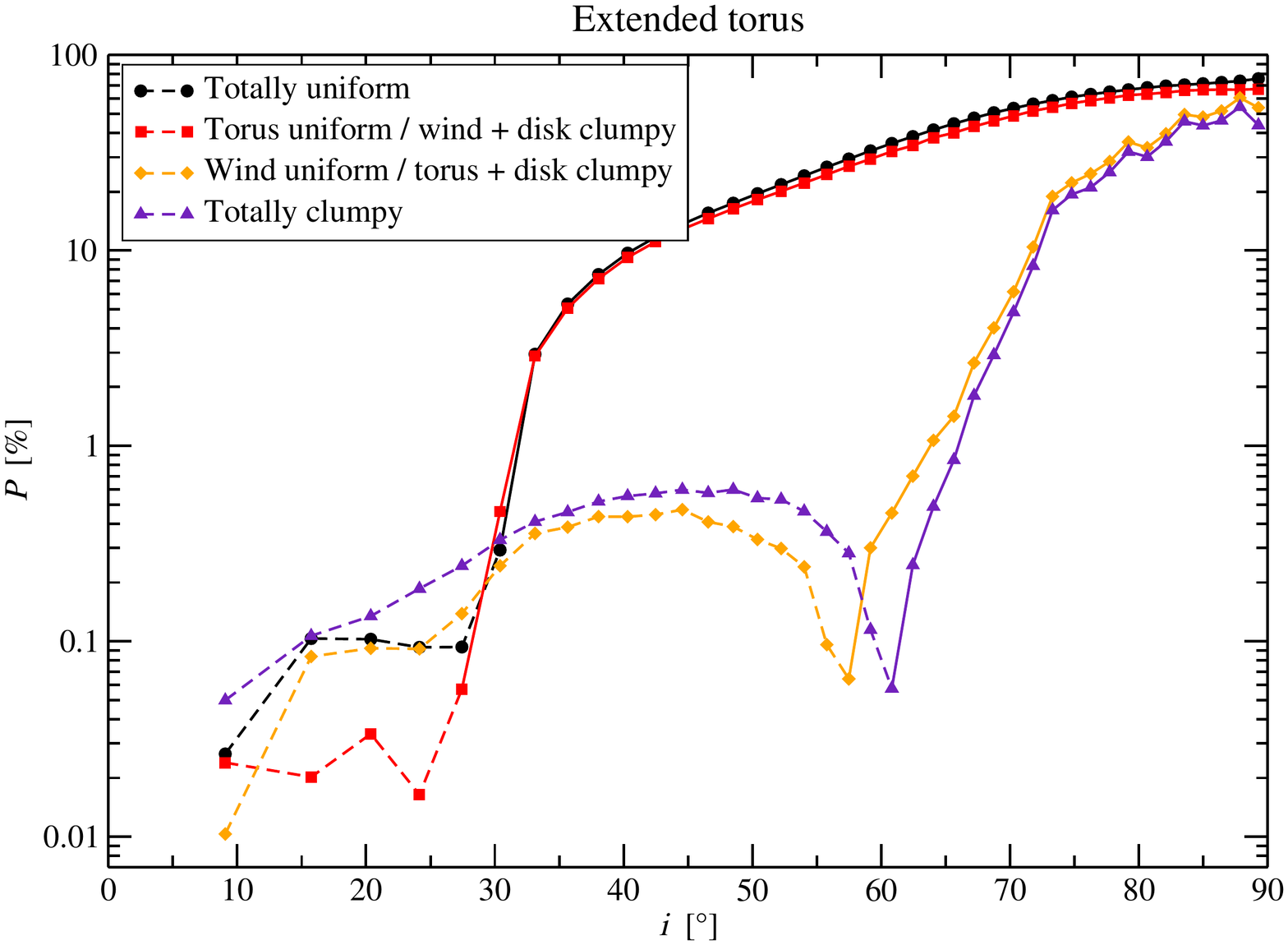}
   \includegraphics[width=9.8cm]{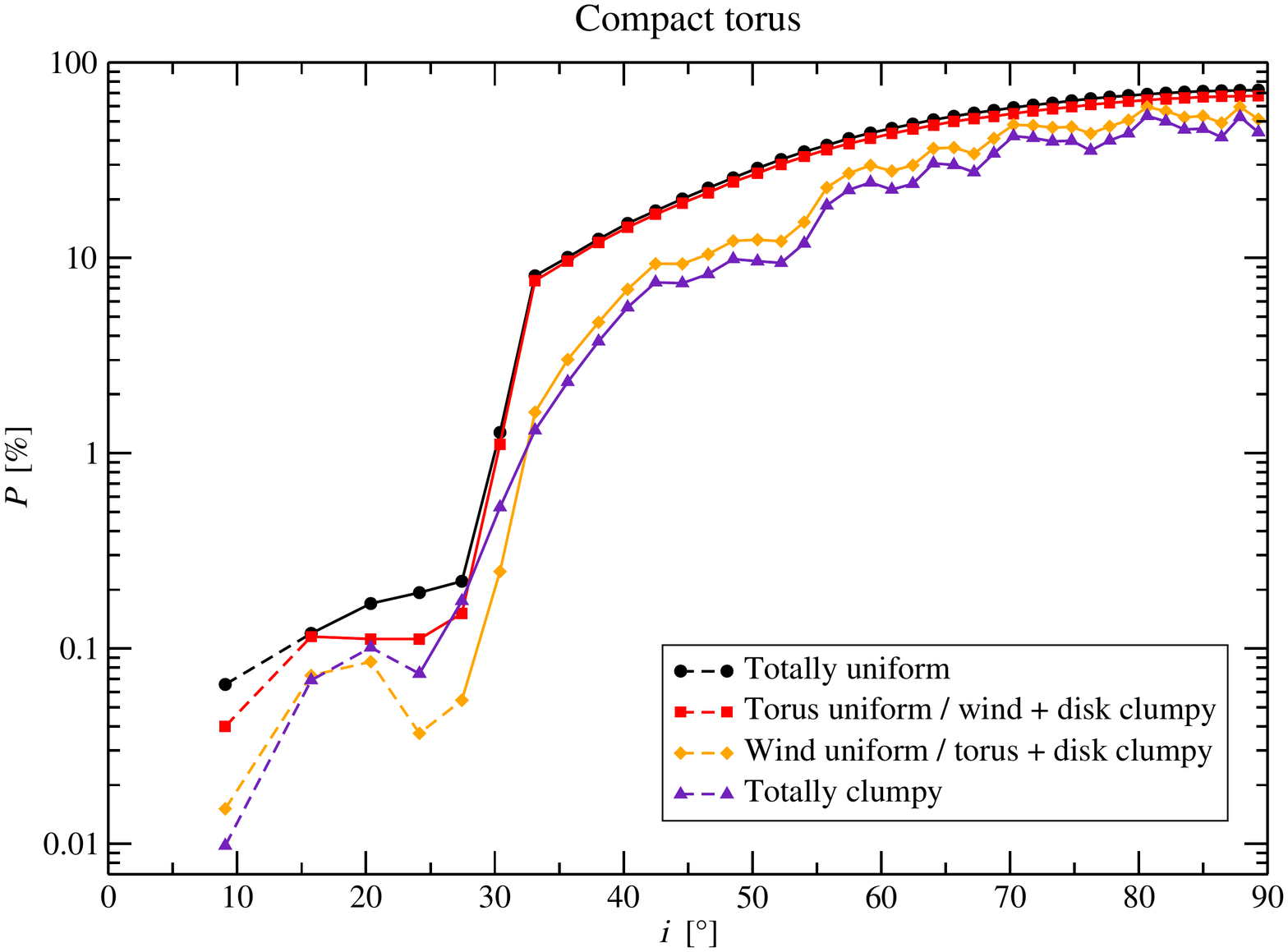}
      \caption{Investigating the impact onto \textit{P} of uniform-density 
	       reprocessing regions within a fragmented model. 
	       A purely uniform-density model is shown with black dots,
	       a clumpy model with a uniform-density dusty torus with red squares,
	       a clumpy model with uniform-density polar winds with orange diamonds
	       and a purely clumpy model with indigo triangles. A dashed line 
	       indicates a polarization position angle $\gamma$~=~90$^\circ$ 
	       (parallel), and a solid line stands for $\gamma$~=~0$^\circ$ 
	       (perpendicular). Top figure: AGN modeling with extended tori;
	       bottom: AGN modeling with compact tori.}
     \label{Fig:Components}%
   \end{figure}

Considering that $P$ is wavelength-independent for all the previous AGN models, we integrate the polarization percentages over 2000 -- 8000~\AA~and compare 
the uniform-density and fragmented AGN to intermediate models. In Fig.~\ref{Fig:Components}, four morphologically-similar cases are plotted versus inclination
for two different geometries of the torus (extended and compact). In black is the completely uniform-density AGN model shown in Paper~II and Fig.~\ref{Fig:AGN} (left column). 
Red squares represent a model with a uniform-density dusty torus but fragmented accretion flow and ionized outflows. In orange is a model with uniform-density polar winds but 
fragmented equatorial regions. Finally, indigo triangles stand for an AGN model purely composed of small clouds. Fig.~\ref{Fig:Components} (top) presents
AGN modeling with extended tori and Fig.~\ref{Fig:Components} (bottom) AGN with compact tori.

As stated in the previous paragraphs, a uniform-density AGN shows low, parallel polarization degrees at type-1 inclinations and high, perpendicular $P$ at type-2s. 
The transition between the two polarization regimes occurs when the observer's line of sight crosses the dust horizon and is clearly recognizable as the 
polarization degree suddenly drops to very low values. This pronounced dip results from the canceling contribution of parallel photons originating from scattering off 
the equatorial disk (and off the inner torus surfaces in the case of extended dusty tori) and perpendicular polarization emerging from compact tori and polar winds. 
The resulting polarization drop, already observed in Paper~II, is a consequence of the sharp-edge torus simplification. If the torus stays uniform, fragmenting 
the polar outflows and the accretion flow has virtually no impact, regardless of the torus geometry. Type-1 polarization remains similar while type-2 polarization is 
marginally lower. However, if the torus is replaced by a cloud distribution but the outflows retain their uniformity, the picture changes drastically. 

Let us study AGN models with extended tori first. The angle at which the polarization dichotomy appears is much larger, close to 60$^\circ$, since the line 
of sight towards the inner AGN region is only partially obscured. Type-1 polarization remains lower than 1~\% and increases at type-2 orientations larger 
than 65$^\circ$, up to 50~\%. When considering a purely clumpy model, such as in Fig.~\ref{Fig:AGN} (middle column), parallel polarization up to 2~\% emerges 
from the model at type-1 views, up to $i~\sim$~70$^\circ$, with the consequences described in the beginning of this section. In addition to that, it is 
interesting to note that the shape of the integrated polarization versus AGN inclination is the same from one model to another, with a polarization hump in 
the type-1 domain and a fast rise of $P$ when the polarization position angle switches from parallel to perpendicular.

For AGN models with compact tori, the situation is similar to a uniform-density AGN model since obscuration is more effective. When the torus remains uniform-density 
but the winds and the equatorial scattering disk are clumpy, the resulting inclination-dependent polarization cannot be disentangled from the uniform-density model.
To alter the AGN polarization signature, a fragmented torus is necessary, reducing the overall polarization by a factor 1.5 to 2. However, in contrast to
an AGN model with an extended toroidal structure, the onset of the polarization dichotomy does not vary. Finally, a fully-fragmented AGN produces a
marginally lower $P$ but conserves the principal characteristics of the uniform-density model.

One of the major conclusion emerging from Fig.~\ref{Fig:Components} is that the impact of a fragmented dusty torus is of utmost importance for AGN modeling
if its dimensions along the midplane extend up to hundreds of parsecs. Extended clumpy tori strongly affect the angle at which the polarization dichotomy 
occurs and help to produce parallel polarization at type-1 angles, even if its half-opening angle is considered to be very small ($\theta_0$ = 30$^\circ$).

%

\subsection{Impact of the overall opening angle}
\label{Clumpy:AGN:op_ang}

   \begin{figure}
   \centering
   \includegraphics[width=9.8cm]{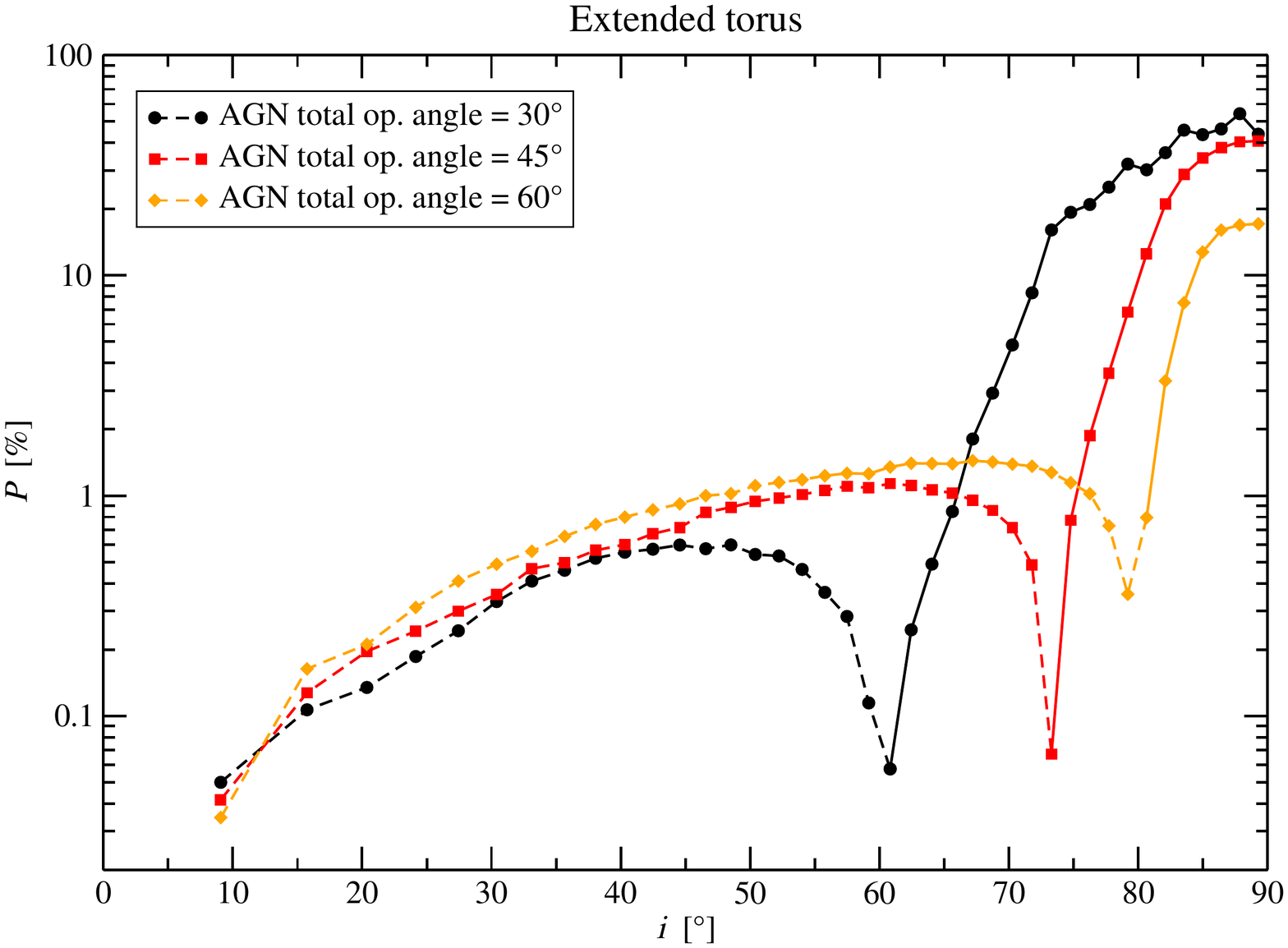}
   \includegraphics[width=9.8cm]{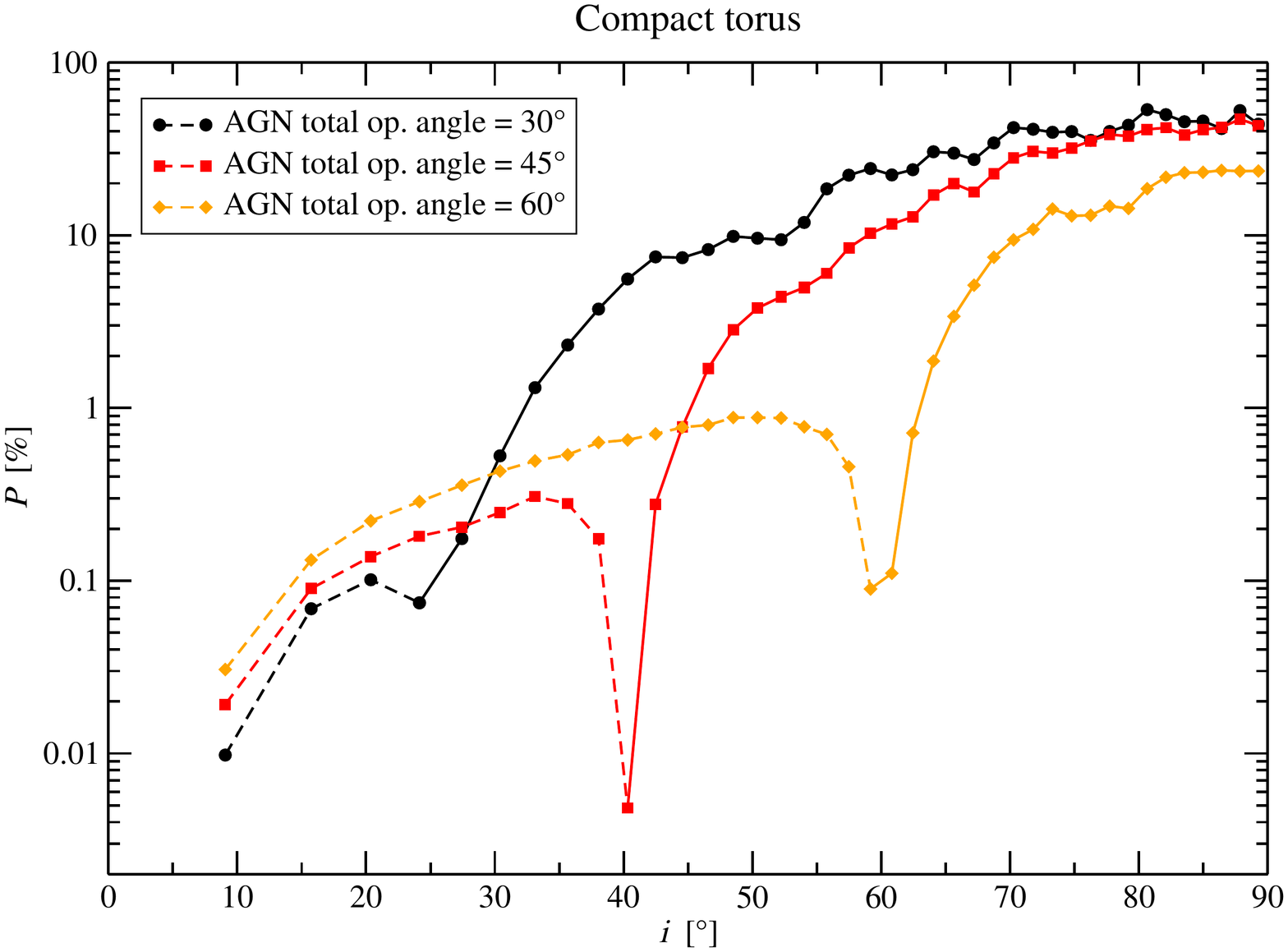}  
      \caption{Investigating the impact of the total
	       opening angle of the system onto the spatially 
	       integrated polarization \textit{P}. A model with total 
	       opening angle of 30$^\circ$ is shown with black dots,
	       45$^\circ$ with red squares and 60$^\circ$ with
	       orange diamonds. All the models are made of fragmented 
	       reprocessing regions (see text). A dashed line 
	       indicates a polarization position angle $\gamma$~=~90$^\circ$ 
	       (parallel), and a solid line stands for $\gamma$~=~0$^\circ$ 
	       (perpendicular). Top figure: AGN modeling with extended, 
	       fragmented tori; bottom: AGN modeling with compact, fragmented tori.}
     \label{Fig:OpAng}%
   \end{figure}
%

It is now clearly identified that most of the differences (in terms of polarization) arise when replacing a uniform, dusty toroidal region with a fragmented 
counterpart. From Paper~II, we know that another impacting parameter is the total opening angle of the model, defined as the half-opening angle 
shared by both the torus and the outflows. To investigate the resulting polarization from various opening angles, we now construct three 
fragmented models. The first one is the clumpy AGN model used in the previous section, with a half-opening angle $\theta_0$ of 30$^\circ$ with respect to the symmetry axis of 
the system (i.e. narrow winds and bulky torus). The second and the third model present larger $\theta_0$, respectively 45$^\circ$ and 60$^\circ$ (wide outflows 
and an oblate torus rather flattened along the equatorial plane). Similarly to Sect.~\ref{Clumpy:AGN:uniform} and Sect.~\ref{Clumpy:AGN:Components}, those 
three models are investigated with either an extended or a compact clumpy torus. The filling factors of the equatorial flow, circumnuclear matter and ionized
winds are the same as the ones used in Sect.~\ref{Clumpy:AGN:uniform}, namely $\cal{F_{\rm flow}} \sim$~0.25, $\cal{F_{\rm torus}} \sim$~0.25, and $\cal{F_{\rm wind}} \sim$~0.06.

Fig.~\ref{Fig:OpAng} (top) shows the inclination-dependent integrated polarimetric signature of the three fragmented AGN models with an extended dusty torus. In comparison 
with the model previously investigated ($\theta_0$ = 30$^\circ$, represented with black dots), a model with $\theta_0$ = 45$^\circ$ (red squares) produces less 
parallel polarization at type-1 angles ($<$~1~\%) since the geometry of the polar winds strengthens the production of diluting, perpendicular polarization.
The turning point between the dominance of parallel and perpendicular polarization starts at $i~\sim$~75$^\circ$, when the inner regions of the model
are effectively hidden behind circumnuclear dust. Perpendicular polarization at type-2 viewing angles rises up to 38~\% at maximum. The third model,
with $\theta_0$ = 60$^\circ$ (orange diamonds) follows a similar trend. The turning point starts at $i~\sim$~80$^\circ$ and a maximal polarization degree
of 16~\% is obtained along equatorial directions. Depending on the parametrization of the model, it is possible to vary both the polarization degree and 
the inclination at which perpendicular polarization dominates. Increasing the number of clumps representative of the polar winds will lead to higher 
perpendicular polarization (see Sect.~\ref{Clumpy:Impact:Winds:Electron} and Fig.~\ref{Fig:Wind_elec}) and thus shift the orientation at which the polarization 
dichotomy appears toward smaller $i$. Nevertheless, we find that extended tori are rather incompatible with observations when the medium is purely clumpy 
(see Sect.~\ref{Clumpy:Discussion}).

Fig.~\ref{Fig:OpAng} (bottom) presents the same analyzes but for AGN models with compact tori. We find that, depending on the global half-opening angle of 
the model, the onset inclination of the polarization dichotomy is very similar to the half-opening angle of the fragmented dusty torus. The net polarization is 
decreasing with flattened tori since their geometry promotes parallel polarization, a behavior already obtained in the case of uniform-density AGN models (Paper~II).
The amount of polarization at type-1 inclinations does not exceed 1\% but for equatorial line of sights $P$ can be as high as 50~\% for tori with a small half opening 
angle, decreasing down to 20~\% for $\theta_0$ = 60$^\circ$. In comparison with AGN models with extended tori, the polarization dichotomy resulting from
AGN with compact tori is more likely to fit polarimetric observations \citep{Marin2014} while reducing the overall polarization degree.

\section{Discussion}
\label{Clumpy:Discussion}

The goal of this paper has been to investigate for the first time the net polarization emerging from non-smooth density distributions of dust grains, and evaluate 
the impact of clumping.

We find that clumping produces several major changes in polarimetric signatures, mainly due to a rather different polarization 
degrees and polarization position angles emerging from clumpy tori. When coupled to other reprocessing regions, the resulting AGN model shows large
deviations from an AGN made of purely uniform-density structures, particularly when extended dusty tori are implemented. It follows that a correct estimation 
of the morphology, filling factor, covering factor and half-opening angle of the torus is necessary to reproduce observed optical/UV polarization measurements. 

In the following, we discuss our modeling results in the framework of past polarimetric observations and recent other numerical simulations.

\subsection{Modeling versus observations}

Modeled under ideal conditions, our AGNs composed of a variety of clumpy reprocessing regions are expected to produce different results according to the 
geometry of their obscuring torus. In the case of extended dusty matter, low polarization degrees ($<$~2~\%) associated with parallel polarization arises 
at type-1 viewing angles, while the transition between parallel and perpendicular polarization occurs at $i~>$~70$^\circ$. Perpendicular $P$ up to 40~\% 
is produced at edge-on inclinations. Using compact tori, AGN models produce lower parallel polarization at type-1 inclinations ($<$~1~\%) and slightly higher
$P$ at type-2s ($\sim$~50~\%) but in our models the transition between type-1 and type-2 classes starts between 24$^\circ$ and 59$^\circ$ (for a 
clumpy torus half opening angle of 30$^\circ$ and 60$^\circ$, respectively). Compared with the AGN models investigated in Paper~II, consisting of 
uniformly-filled scattering regions, fragmentation leads to higher polarization along polar directions and lower $P$ along the equator. Moreover, while a 
uniform-density dusty torus with $\theta_0$ = 60$^\circ$ was necessary to produce a transition between parallel and perpendicular polarization at 
$i~\sim$~60$^\circ$, we have shown in Sect.~\ref{Clumpy:AGN} that extended fragmented tori tend to shift this value to higher inclinations due 
to less effective obscuration along observer's viewing angles grazing the torus horizon.

It is interesting to evaluate the relevance of fully fragmented AGN models within the framework of the compendium of inclinations of Seyfert galaxies published by \citet{Marin2014}.
In this work, 53 estimated AGN inclinations are matched with ultraviolet/optical continuum polarization measurements and constraints for AGN modeling are derived.
The transition between type-1 and type-2 inclination is found to occur between 45$^\circ$ and 60$^\circ$, which is in agreement with our modeling of a
clumpy AGN with a compact dusty torus and global half-opening angles of 30$^\circ$ to 60$^\circ$. The level of parallel polarization at type-1 orientations is 
correctly reproduced (except for several enigmatic, highly polarized type-1 AGN, see below), as well as the amount of parallel polarization at type-2 inclinations. 
An AGN model with a compact, but disrupted, dusty torus is thus in agreement with observations and helps to decrease the production of perpendicular polarization at 
type-2 inclinations. However, AGN models with extended clumpy tori cannot reproduce the observed polarization dichotomy onset ($<$~60$^\circ$) since 
their oblate morphology reduces the efficiency of obscuration at intermediate angles for a moderate filling factor. This means that extended tori are incompatible 
with observations\footnote{This is in agreement with the X-ray results obtained by \citet{Gandhi2015}, who shown that the observed Fe K$\alpha$ emission 
radii of 9 type-1 AGN are similar to, or smaller than the radius of the BLR, favoring compact tori.} and future simulations must focus on compact structures, 
independently of their fragmentation level. 

Several issues remain. Results from \citet{Marin2014} show that $P$ is expected to rise promptly, from 0.7~\% to 30~\%, between inclinations of 45$^\circ$ and 60$^\circ$ 
a behavior that cannot be reproduced by a circumnuclear distribution of dust clouds. In addition to that, the question of highly polarized ($P~>$~4~\%) type-1 AGN,
such as Fairall~51 \citep{Martin1983}, IC~4329A \citep{Brindle1990} or ESO~323-G077 \citep{Schmid2003} holds still. Our analysis confirms that clumpy AGNs with compact tori are 
not the solution to reproduce high parallel or perpendicular polarization at type-1 inclinations and that another solution must be found. It is clear from our 
series of papers that the unified model of AGN, established from polarimetric observations, is true at zeroth order but detailed radiative transfer calculations
point towards a need for adjustments. Theoretical models treating the reprocessing regions of AGN as dynamical structures started to emerge in the 
beginning of the millennium with the work of \citet{Proga1998,Proga1999}, based on similar studies for cataclysmic variables and OB stars (e.g. \citealt{Pereyra1997}).
At the same time, \citet{Elvis2000} suggested a phenomenological model for quasar disk winds. Those models predict that mass-loss from AGN 
is in the form of radiation-driven outflows, where momentum is extracted from the radiation field near accreting disks by line opacity \citep{Castor1975}.
Such winds could be associated with the AGN ionization cones and also be responsible for a failed dusty wind along the equator \citep{Elitzur2006,Gallagher2014}, where dust can form and survive 
if shielded from the full continuum by inner highly-ionized disk-born winds \citep{Elvis2012,Czerny2012,Czerny2014}. Such models are being currently investigated in order to 
see if they can reproduce the observed spectroscopic and polarimetric signatures of AGNs. Preliminary studies based on the model of \citet{Elvis2000} showed that 
line-driven accretion disk wind models are good candidates to replace the usual axisymmetric unified model \citep{Marin2013b,Marin2013c,Marin2013d,Marin2014}.

\subsection{Clumpy tori and ``changing look'' AGN}

In the X-ray domain, several intermediate type AGN (1.5, 1.8, 1.9) have been discovered displaying rapid transitions between reflection-dominated (Compton thick, 
as in our models) to transmission-dominated (Compton-thin) states. Such dramatic X-ray spectral changes are characteristic of a sub-class of Seyfert galaxies 
called ``changing look`` AGN \citep{Risaliti2005}. Those variations are attributed to clouds composed of optically thick, cold atomic material transiting in 
front of the observer's line of sight, creating X-ray eclipses, rather than to intrinsic emission variability. The rapid time-scales of those eclipses,  
lasting from hours to days \citep[e.g.][]{Elvis2004}, suggest that the absorbers are situated on compact scales close to the low-ionization-line broad line region 
(LIL BLR, see review by \citealt{Turner2009}). Longer eclipses, up to months, have also been detected in 8 objects by \citet{Markowitz2014}, indicating a possible 
location of the obscuring cloud within the inner regions of infrared-emitting dusty tori.

The analogy between ``changing look'' AGN and clumpy tori is straightforward. If the molecular clouds composing a fragmented torus are in a dynamical 
state, as expected by \citet{Krolik1988}, Seyfert-galaxies within a specific range of inclinations should show variation in their 
spectroscopic and polarimetric properties. Since the apparent motion of the torus gas clouds is much slower than the clouds from the BLR 
\citep{Risaliti2002}, long-term monitoring is necessary to investigate these potential phenomena. By extension, our paper shows that we expect differences 
in terms of polarization between two AGNs with the same inclination and a clumpy distribution of absorbing matter, but with the line of sight of the 
first model being obscured by a clump while the other model is free of absorption. From Fig.~\ref{Fig:Components}, we see that for an inclination of 
50$^\circ$ and an extended torus, a perpendicular polarization of about 50~\% is expected when the central engine is obscured while, if there is no 
clump along the observer's view, parallel polarization with $P~<$~2~\% is more probable. In the case of a compact yet fragmented torus, $P$ would 
decrease from 25~\% to 9~\% for the same inclination. The real change in polarization is likely to be less dramatic since neighboring clumps would 
also affect the net polarization degree, but one can expect a variation of several percent in the polarization and a possible switch of the position 
angle. Preliminary spectropolarimetric modeling has been conducted in the X-ray range for a clumpy distribution of gas clouds around the central part 
of NGC~1365 by \citet{Marin2013} and a switch of the polarization position angle has been predicted between the soft and the hard X-ray bands.

\subsection{An alternative solution to fragmentation}

In Paper~I and Paper~II, we considered smooth dusty tori (i.e., continuous dust distributions that change either slowly or not at all with distance from the 
AGN engine). In this paper, we have investigated the case of circumnuclear obscuration using a fragmented medium composed of constant density spheres. Both 
models have failures and successes in reproducing peculiar aspects of AGN observations (see also the discussions in \citealt{Vollmer2004,Dullemond2005,Feltre2012}).

However, there is a third, less known competitor: bulk tori maintained by nuclear starbursts. In this scenario, supernovae (SNe), from nuclear star-bursts 
situated within the inner 100~pc of low-luminosity AGNs (LLAGNs, $L_{\rm bol}~\leqslant$~10$^{42}$~ergs.s$^{-1}$), could put out enough energy to keep up the scale height of 
the dusty torus \citep{Fabian1998}. The bulky torus is essentially supported from the inside by SNe energy boosts and is then neither smooth nor fragmented, but has a 
``Swiss cheese'' or ``sponge-like'' structure. The proposed scenario of \citet{Fabian1998} was investigated in detail by \citet{Wada2002} using three-dimensional 
hydrodynamical simulations. The authors assumed a toroidal geometry around the central supermassive black hole (SMBH) in which the energy input from SNe is in 
equilibrium with the turbulent energy dissipation. Doing so, they infer new constraints on the inclination of LLAGNs as the torus half-opening angle seems to be 
correlated with the SMBH mass. The polarization signal of such a scenario remains unexplored and could lead to different signatures than the uniform-density and 
the fragmented tori. But one must keep in mind that, according to \citet{Krolik1988}, stirring by stellar processes might not be strong enough to compete with 
energy dissipation in the circumnuclear gas.

\section{Conclusions and future work}
\label{Clumpy:Conclusion}

We have modeled a set of isolated clumpy structures and have run Monte Carlo simulations in order to compare their resulting polarization signatures with the uniform
models analyzed in Paper~II. We have created three different fragmented models using spheres of constant radius and density, corresponding to the respective 
filling factors: $\cal{F} \sim$~0.06, $\cal{F} \sim$~0.13 and $\cal{F} \sim$~0.25. 

We find that equatorial distributions of clumps, either for a dusty torus or an ionized accretion flow, tend to decrease the net polarization percentage,
independently of the filling factor. The resulting polarization position angle depends on the half-opening angle and on the morphology of the given region. Extended 
clumpy configurations with half-opening angles $<$~=~45$^\circ$ give rise to parallel polarization for type-1 and type-2 viewing angles, while similar compact
yet fragmented tori produce perpendicular polarization. A residual photon flux always manages to escape from the central parts of the model even when considering an 
observer's line of sight passing through the highly-covered equatorial region. Due to multiple scattering and thus depolarization, this equatorial flux is carrying 
a weak polarization degree ($<$~1~\%) that is likely to be diluted by star-light emission. In the case of scattering regions distributed along the polar directions, 
the impact of fragmentation is reduced: a small number of clouds is able to properly reproduce the polarization behavior of uniform-density models. Increasing the 
cloud distribution only results in increasing the net polarization percentage as the total optical depth increases along the observer's line of sight. As in the 
uniform-density density case, fragmented outflows solely produce perpendicular polarization angles.

Combining the radiation supported electron region with the dusty torus and the polar outflows, we then modeled AGNs only made of clumpy structures. 
Such models, with a torus half-opening angle of 30$^\circ$, produce different results according to the radial size of the clumpy tori. For extended tori (wider 
than tens of parsecs) a small polarization percentage (up to 2~\%), associated with parallel polarization, arises at inclinations up to $i~\sim$~70$^\circ$. 
When the observer's line of sight passes through the central part of the equatorial dusty matter, polarization increases and switches from parallel to perpendicular 
orientation. A large $P$ ($\sim$ 40~\%) is found at edge-on viewing angles. AGN models with compact tori behave rather similarly to uniform-density models except that
they produce significantly lower polarization at intermediate and type-2 orientations. The observed polarization dichotomy is better reproduced 
by models with compact yet fragmented tori with $\theta_0~\ge$~45$^\circ$.

It was also shown that fragmented dusty tori have the strongest impact on $P$. Large, fragmented tori are not viable solutions to replace geometrically flat, 
uniform-density dust distributions regardless of their half-opening angle; compact circumnuclear structures must be favored. Hence, particular attention should 
be focused on the circumnuclear region: while unfragmented tori give similar results in terms of intensity fluxes and polarization behavior, cloudlet distributions 
of absorbing gas that are able to reproduce the observed spectroscopic signatures of nearby Seyfert galaxies might not work when polarization is considered. 
To check the validity of compact, clumpy AGN tori, multi-wavelength and time-resolved spectropolarimetric modeling are needed. We are planning to undertake this 
in future work.

\acknowledgements We thank the anonymous referee who provided insight and expertise that 
greatly assisted our paper. This research has been supported by the Academy of Sciences of the 
Czech Republic, the French PNHE and the grant ANR-11-JS56-013-01 ``POLIOPTIX''. Part of this 
work was supported by the COST Action MP1104 ``Polarization as a tool to study the Solar System 
and beyond''.

\bibliographystyle{aa}
\bibliography{biblio}

\end{document}